\PassOptionsToPackage{unicode}{hyperref}
\PassOptionsToPackage{hyphens}{url}
\documentclass[
  12pt,
]{article}
\usepackage{amsmath,amssymb}
\usepackage{iftex}
\ifPDFTeX
  \usepackage[T1]{fontenc}
  \usepackage[utf8]{inputenc}
  \usepackage{textcomp} 
\else 
  \usepackage{unicode-math} 
  \defaultfontfeatures{Scale=MatchLowercase}
  \defaultfontfeatures[\rmfamily]{Ligatures=TeX,Scale=1}
\fi
\usepackage{lmodern}
\ifPDFTeX\else
\fi
\IfFileExists{upquote.sty}{\usepackage{upquote}}{}
\IfFileExists{microtype.sty}{
  \usepackage[]{microtype}
  \UseMicrotypeSet[protrusion]{basicmath} 
}{}
\makeatletter
\@ifundefined{KOMAClassName}{
  \IfFileExists{parskip.sty}{%
    \usepackage{parskip}
  }{
    \setlength{\parindent}{0pt}
    \setlength{\parskip}{6pt plus 2pt minus 1pt}}
}{
  \KOMAoptions{parskip=half}}
\makeatother
\usepackage{xcolor}
\usepackage[margin=1in]{geometry}
\usepackage{graphicx}
\makeatletter
\def\maxwidth{\ifdim\Gin@nat@width>\linewidth\linewidth\else\Gin@nat@width\fi}
\def\maxheight{\ifdim\Gin@nat@height>\textheight\textheight\else\Gin@nat@height\fi}
\makeatother
\setkeys{Gin}{width=\maxwidth,height=\maxheight,keepaspectratio}
\makeatletter
\def\fps@figure{htbp}
\makeatother
\usepackage{url}
\usepackage{setspace}
\usepackage[belowskip=0pt,aboveskip=0pt]{caption}

\usepackage{xr}
\usepackage{xspace}
\usepackage{rotating}
\usepackage{bm}
\usepackage{float}
\usepackage{caption,graphics}
\usepackage{graphicx}
\usepackage{makecell}
\usepackage{xr}
\usepackage{enumerate}
\usepackage{ifthen}
\newboolean{showsupplement}
\setboolean{showsupplement}{false}

\renewcommand{\arraystretch}{1.5}

\usepackage{float} 

\usepackage{booktabs}
\usepackage{longtable}
\usepackage{array}
\usepackage{multirow}
\usepackage{wrapfig}
\usepackage{float}
\usepackage{colortbl}
\usepackage{pdflscape}
\usepackage{tabu}
\usepackage{threeparttable}
\usepackage{threeparttablex}
\usepackage[normalem]{ulem}
\usepackage{makecell}
\usepackage{xcolor}
\ifLuaTeX
  \usepackage{selnolig}  
\fi
\usepackage[]{natbib}
\usepackage{bookmark}
\IfFileExists{xurl.sty}{\usepackage{xurl}}{} 
\hypersetup{
  pdftitle={Enhancing data-limited assessments with random effects: A case study on Korea chub mackerel (Scomber japonicus)},
  pdfauthor={Kyuhan Kim1,; Nokuthaba Sibanda2,; Richard Arnold2,; Teresa A'mar1,3},
  hidelinks,
  pdfcreator={LaTeX via pandoc}}

\title{Enhancing data-limited assessments with random effects: A case
study on Korea chub mackerel (\emph{Scomber japonicus})}
\author{Kyuhan Kim\textsuperscript{1}\textsuperscript{*}, \and Nokuthaba
Sibanda\textsuperscript{2}, \and Richard
Arnold\textsuperscript{2}, \and Teresa A'mar\textsuperscript{1,3}}
\date{}

\begin{document}
\maketitle

\(\textsuperscript{*}\)Corresponding author:
\href{mailto:kh2064@gmail.com}{\nolinkurl{kh2064@gmail.com}}\\
\(^1\)Dragonfly Data Science, PO Box 27535, Wellington 6141, New
Zealand\\
\(^2\)School of Mathematics and Statistics, Victoria University of
Wellington, PO Box 600, Wellington 6140, New Zealand\\
\(^3\)Current affiliation: Meridian Energy Limited, PO Box 10840,
Wellington 6143, New Zealand\\

\pagebreak

\section*{Abstract}\label{abstract}
\addcontentsline{toc}{section}{Abstract}

In a state-space framework, temporal variations in fishery-dependent
processes can be modeled as random effects. This modeling flexibility
makes state-space models (SSMs) powerful tools for data-limited
assessments. Though SSMs enable the model-based inference of the
unobserved processes, their flexibility can lead to overfitting and
non-identifiability issues. To address these challenges, we developed a
suite of state-space length-based age-structured models and applied them
to the Korean chub mackerel (\textit{Scomber japonicus}) stock. Our
research demonstrated that incorporating temporal variations in
fishery-dependent processes can rectify model mis-specification but may
compromise robustness, which can be diagnosed through a series of model
checking processes. To tackle non-identifiability, we used a
non-degenerate estimator, implementing a gamma distribution as a penalty
for the standard deviation parameters of observation errors. This
penalty function enabled the simultaneous estimation of both process and
observation error variances with minimal bias, a notably challenging
task in SSMs. These results highlight the importance of model checking
and the effectiveness of the penalized approach in estimating SSMs.
Additionally, we discussed novel assessment outcomes for the mackerel
stock.

\textbf{Keywords}: Chub mackerel, CPUE standardization, random effect,
stock assessment, state-space model

\pagebreak

\section{Introduction}\label{introduction}

\subsection{Trade-off between model complexity and data
availability}\label{trade-off-between-model-complexity-and-data-availability}

Fisheries management faces many challenges, especially when dealing with
fish stocks that have limited data available. In these cases, we often
rely on fishery-dependent information, primarily catch-per-unit-effort
(CPUE) data, as our primary source of information
\citep{maunder2004standardizing, maunder2006interpreting, hoyle2024catch}.
However, CPUE is inherently subject to various influences, such as
fluctuations in fishing efficiency, variations in environmental
conditions, changes in fish availability to fishing gear, and shifts in
the geographic scope of fishing grounds
\citep{maunder2004standardizing, maunder2006interpreting, wilberg2009incorporating, hoyle2024catch}.
These inherent variations in CPUE render it unreliable as an index of
abundance without proper adjustments
\citep{maunder2004standardizing, maunder2006interpreting, hoyle2024catch}.

As a result, CPUE standardization has become standard practice in
fisheries stock assessment. Over time, several approaches have been
developed for standardizing CPUE data, including generalized linear
models, generalized additive models, and generalized linear mixed models
\citep{maunder2004standardizing, wilberg2009incorporating, hoyle2024catch}.
However, many of these methods require supplementary data, including
information about when and where fishing activities occur, vessel
specifications, environmental conditions, and the composition of the
target species. Unfortunately, many fish stocks, particularly those
newly brought under management, lack this supplementary information,
making the application of standardization methods impractical for these
stocks. Furthermore, these standardization models can only account for
explicitly included factors, potentially leaving CPUE data vulnerable to
unexplained influences \citep{wilberg2009incorporating}.

In situations where auxiliary information for standardization is
unavailable, an alternative approach to address unexplained variation in
CPUE is to employ CPUE standardization models within a stock assessment
framework. This approach is similar to the methods described in
\citet{maunder2001general} and \citet{maunder2004integrating}, but with
a shift toward using descriptive models instead of functional models.
These descriptive models allow for temporal variation without explicitly
specifying the underlying mechanism
\citep{wilberg2006performance, wilberg2009incorporating}. While this
approach involves estimating additional parameters, potentially leading
to challenges such as over-parameterization and non-identifiability
\citep{auger2016state, auger2021guide, hyun2022evaluation, kim2022state},
adopting a state-space framework where temporal variations are treated
as random effects
\citep{nielsen2014estimation, miller2016state, miller2018evaluating, trijoulet2020performance, stock2021woods, kim2022state}
can help mitigate these issues \citep{punt2023those}.

In a state-space framework, unexplained variation in CPUE can be modeled
as random effects. The relationship between CPUE and actual fish
abundance is generally influenced by fishery-dependent factors such as
selectivity (including components such as availability and contact
selectivity) and fishing efficiency, which are not time-invariant in
practice \citep{hoyle2024catch}. Thus, the investigation of time-varying
selectivity and catchability within assessment models has been explored
in previous studies
\citep{wilberg2009incorporating, martell2014towards, punt2014model}.
These studies suggested that integrating time-varying selectivity and
catchability into assessment models can enhance performance and
alleviate retrospective patterns --- systematic changes in population
size estimates or other derived quantities resulting from the addition
or removal of recent data points
\citep{wilberg2006performance, martell2014towards, hurtado2015looking}.
However, this approach, while more parsimonious than conventional
fixed-effects models, introduces additional model parameters,
potentially leading to robustness issues \citep{deroba2015simulation}.

Addressing time-varying factors through random effects requires
estimating multiple variance parameters, a task complicated by
non-identifiability issues that can arise in state-space models
\citep{auger2016state, auger2021guide}. Even simple models encounter
this problem, as estimates often approach zero when attempting to
estimate both observation and process error variance parameters
\citep{auger2016state, hyun2022evaluation, kim2022state}. This issue
becomes more pronounced in assessment models with multiple variance
parameters \citep{kim2022state}. To confront this challenge, previous
studies have employed various strategies, such as incorporating
observation error variance as an input
\citep{miller2016state, miller2018evaluating, trijoulet2020performance},
assuming that observation and process error variances are equal
\citep{ono2012model, parent2012introduction, thorson2015mixed, rankin2015alternative},
or imposing prior distributions on the observation and process error
variance parameters
\citep{pedersen2017stochastic, stock2021woods, ICES2023}. However, these
approaches may yield biased estimates if the assumed values are
inaccurate or if reliable prior information is unavailable. A potential
solution is to use a non-degenerate estimator, which involves maximizing
the marginal likelihood with a gamma distribution as a penalty function
for parameters with boundary estimates, similar to applying a weakly
informative gamma prior in Bayesian analysis
\citep{chung2013nondegenerate, gelman2020bayesian}. While this approach
is recommended in the Bayesian framework \citep{gelman2020bayesian}, its
practical performance in fisheries stock assessments remains unexplored,
raising questions about its applicability and efficacy.

\subsection{Main objectives of this
study}\label{main-objectives-of-this-study}

In this study, we developed a length-based age-structured model in a
state-space framework, where time-varying selectivity and catchability
were incorporated. The model was designed to address unexplained
temporal variability in CPUE when additional information for
standardization is unavailable. We applied the model to the Korean
mackerel (\textit{Scomber japonicus}) stock as a case study. We chose
the mackerel stock as a case study because the stock has been influenced
by various fishery-dependent factors, including changes in fishing
efficiency and fishing grounds \citep{lee2011variation, seo2017change}.
One potential issue with the mackerel stock is that CPUE has increased
since 1985, but the reasons for this increase are not well understood
and have not been accounted for in stock assessments. Without a clear
understanding of the factors driving the increase in CPUE, the common
assumption that CPUE is proportional to abundance may not hold,
potentially leading to overly optimistic stock assessment results. To
the best of our knowledge, detailed information on fishery behavior for
this mackerel stock, essential for CPUE data standardization, has not
been recorded or is not in an accessible form. This absence of
information is evident, as all recent published quantitative assessments
of this stock have used the CPUE data without standardization
\citep{jung2019bayesian, gim2019length, gim2020management, jung2021bayesian, hong2022stock, gim2022application, kim2022state, gim2023assumptions}.
Despite the incomplete understanding of the influence of
fishery-dependent factors on the assessment results, the stock has been
managed using a quota-based system since 1999
\citep{sim2020analysis, hyun2023overview, fira_tac2023}.

Such limited understanding and information regarding fishery-dependent
processes in the mackerel stock pose a significant challenge in
providing robust and reliable stock assessment estimates. This challenge
raises critical questions about the scientific foundation of the quota
management system in Korea \citep{hyun2023overview}. Our primary
objective is to investigate potential time-varying components of
fishery-dependent factors, specifically catchability and selectivity,
under the data-limited conditions that the mackerel stock is currently
facing. Our focus is on maintaining the statistical robustness of the
model in terms of parameter identifiability (while estimating both
process and observation variance parameters, which is often problematic
in state-space models, as discussed above) and the validity of model
assumptions while keeping the model as realistic as possible to reflect
the time-varying nature of fishery-dependent factors. Additionally, we
aim to understand how these time-varying components influence assessment
results.

To validate the effectiveness of our proposed model, we conducted a
comprehensive evaluation under various assumptions related to
selectivity and catchability. Through an extensive simulation study, we
assessed the model's performance, examining aspects such as robustness,
uncertainties, and biases. To validate assumptions associated with
random effects, we examined goodness-of-fit of the model using a
residual analysis. Moreover, we compared the model's performance with
and without the non-degenerate estimator. This comparative analysis
allowed us to investigate the efficacy of the penalized likelihood
approach in addressing non-identifiability issues within state-space
stock assessment models.

\section{Materials and methods}\label{materials-and-methods}

In this section, we describe the data and methods used in this study. We
first provide background information on the chub mackerel stock in
Korean waters for better understanding of the data and model structure.
We then describe the data used in this study, followed by the model
structure, estimation procedure, and model checking processes. The
notation used in this study is summarized in Table \ref{table:terms}.

\subsection{Background information on the mackerel
stock}\label{background-information-on-the-mackerel-stock}

Government-issued quotas for the total allowable catch (TAC) have
regulated the chub mackerel stock in Korean waters, with a specific
focus on the large purse seine (LPS) fishery since 1999
\citep{fira_tac2023}. Despite historical records indicating the presence
of more than six different fisheries targeting the mackerel stock (e.g.,
large purse seine, small purse seine, gillnet, trawl, etc.), the LPS
fishery alone contributes to over 90\% of the annual total catch of chub
mackerel in Korea \citep{jung2019bayesian}.

The Korean National Institute of Fisheries Science (NIFS) has been
collecting CPUE data since 1976 \citep{jung2019bayesian}. For the
computation of CPUE, NIFS collected data on catch (in weight) and effort
from approximately 70\% of all fishing trips undertaken by the LPS
fishery annually. The harvesters involved in these trips were
responsible for providing their catch and effort, with the latter
measured in terms of the number of hauls \citep{jung2019bayesian}.

In addition to the CPUE data, NIFS has collected samples of fish from
landed catches by the LPS fishery to obtain biological parameters,
including length, weight, and sexual maturity \citep{kim2020maturity}.
While length and weight of the samples have been periodically measured,
age determination has been rarely conducted
\citep{kang2015first, jung2021study}. Length frequency data from the LPS
fishery have also been collected since 2000
\citep{kim2018inference, gim2019length}.

To establish a scientific foundation for the TAC-based management of
this stock, numerous prior studies have evaluated the mackerel stock
using the aforementioned fishery-dependent data. However, all these
previous studies made the assumption that the nominal CPUE data were
directly proportional to the abundance of the stock
\citep{jung2019bayesian, gim2019length, gim2019importance, gim2020management, jung2021bayesian, hong2022stock, gim2022application, kim2022state},
relying on a single constant catchability parameter. Furthermore, when
structured models were employed, it was assumed that selectivity
remained constant over time
\citep{gim2019length, gim2019importance, gim2020management, gim2022application, kim2022state}.

These assumptions persisted, despite earlier studies discussing the
temporal increase in vessel power in the LPS fishery
\citep{seo2017change} and temporal changes in its fishing locations
\citep{lee2011variation}. Ignoring these factors in CPUE data used for
stock assessments could result in biased estimates of stock abundance
due to potential hyperstability \citep{hilborn1992quantitative}.

\subsection{Data}\label{data}

\begin{figure}

{\centering \includegraphics[width=0.95\linewidth]{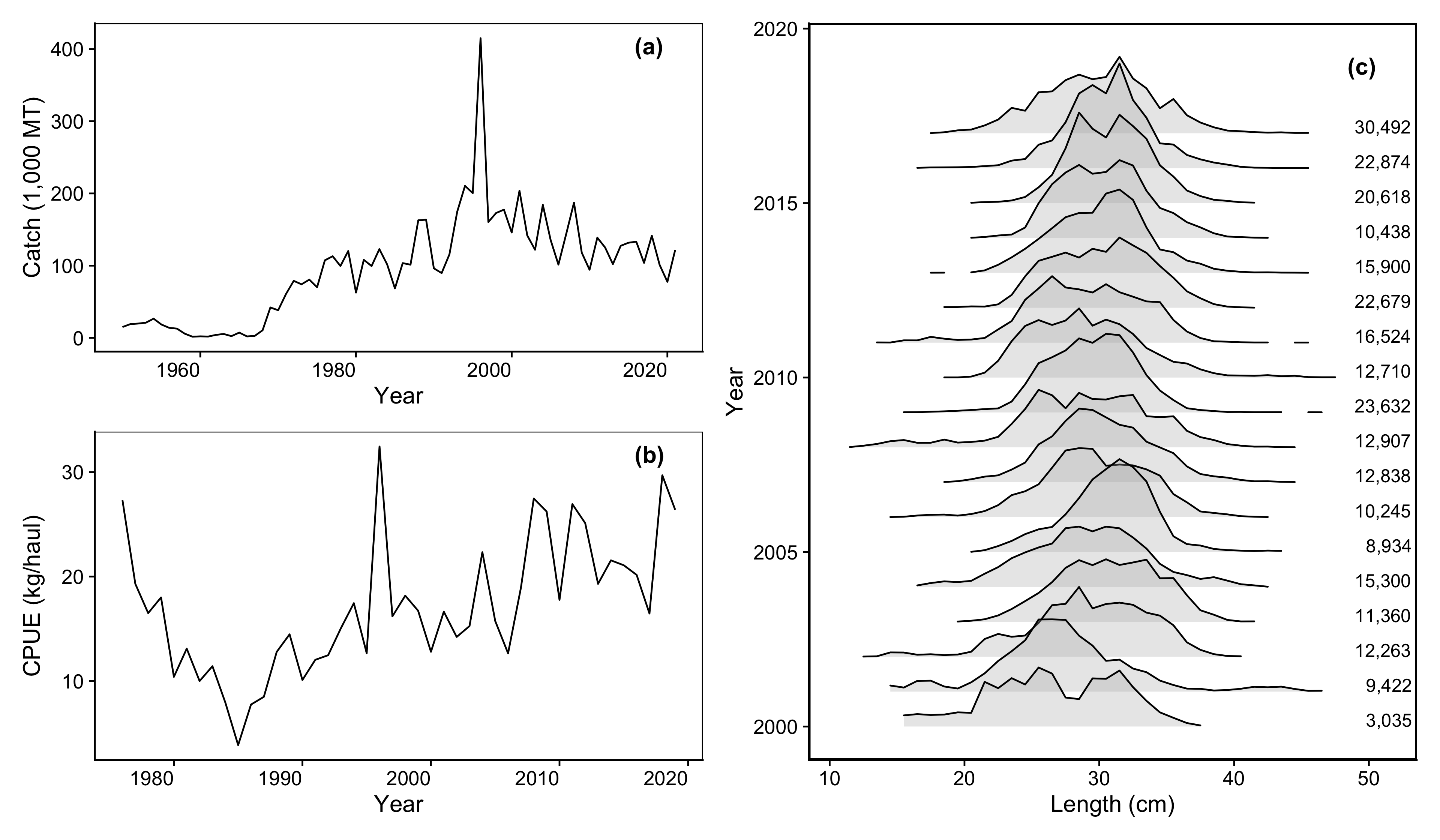} 

}

\caption{Catch (a), catch-per-unit-effort (CPUE) (b), and length composition (c) data from the mackerel stock. The CPUE and length composition data were collected from the large purse seine fishery only. The length composition data are shown as the proportion of fish in each length bin. The numbers in the panel (c) indicate sample sizes for each year. \label{MackData}}\label{fig:unnamed-chunk-13}
\end{figure}

As in previous studies that used structured population dynamics models
\citep{gim2020management, gim2022application, kim2022state, gim2023assumptions},
this research utilized three datasets: CPUE data from 1976 to 2019,
length composition data from 2000 to 2017, and annual catch (in weight)
data from 1950 to 2021 (Fig. \ref{MackData}). The CPUE and length
composition data were not available in a numerical form, so we extracted
them from earlier publicly accessible studies
\citep{kim2018inference, jung2019bayesian, gim2019length} by digitizing
their published figures using WebPlotDigitizer (a process often referred
to as reverse engineering) \citep{Rohatgi2022}. The accuracy of the
digitized data was verified by comparing them with the original figures.
Annual catch data were sourced from Statistics Korea (1976 to 2021) and
the FishStatJ database (1950 to 1975) provided by the Food and
Agriculture Organization of the United Nations (FAO)
\citep{fishstatj2020fishstatj}.

\subsection{Process models}\label{process-models}

We developed a length-based age-structured model as the primary process
model to estimate the abundance of the chub mackerel stock. Within this
model, we explored several descriptive models for catchability and
selectivity. Our approach is akin to previous studies by
\citet{wilberg2006performance} and \citet{wilberg2009incorporating},
where different modeling approaches for time-varying catchability were
tested. However, we extended this approach to include time-varying
selectivity as well.

This methodology allowed us to investigate potential time-varying
components of these two fishery-dependent factors without requiring
additional information on the underlying mechanisms. Further details on
each model component are provided in the subsequent sections.

\subsubsection{Population dynamics}\label{population-dynamics}

The estimation of abundance for ages and years was conducted through an
age-structured population dynamics model. Within this model, annual
recruitment was assumed to be independent of the spawning stock biomass
(SSB) as in previous studies
\citep{szuwalski2015examining, hilborn2017does, miller2018evaluating}.
This assumption was also supported by \citet{hiyama2002stock}, who found
no significant correlation between annual recruitment and SSB of chub
mackerel in the adjacent waters of Japan.

We assumed a recruitment age of 0, as suggested by the catch-at-age
estimates presented in \cite{jung2021study}. In their study, the authors
developed an age-length key table using catch samples from the LPS
fishery in 2015. This table was used to categorize the 2015 catch into
distinct age classes. However, it should be noted that we did not
incorporate this single age-length key table into our study. This
decision was due to its year-specific nature, with detailed explanations
provided in the \textit{Length-at-age distributions in catch} section
below.

The natural logarithm of the annual recruitment \(\log(N_{0,t})\) was
assumed to follow a normal distribution with a mean of \(\log(R)\) and a
variance of \(\sigma^{2}_{R}\). Thus, the annual abundance of age \(a\)
in year \(t\) was calculated as follows:

\begin{equation*}
\begin{aligned}
N_{a,t}=
\begin{cases}
R \cdot e^{\varepsilon^{R}_{t}}, & \text{ for } a=0 \\  
N_{a-1,t-1} \cdot e^{-Z_{a-1,t-1}}, & \text{ for } 0<a<A \\ 
N_{a-1,t-1} \cdot e^{-Z_{a-1,t-1}}+N_{a,t-1} \cdot e^{-Z_{a,t-1}}, & \text{ for } a=A \\
\end{cases}.
\end{aligned} 
\end{equation*}

Here, subscripts \(a\) and \(t\) represent the indices for age and year,
respectively. \(N_{a,t}\) represents the abundance (in numbers) of age
\(a\) in year \(t\). \(\varepsilon^{R}_{t}\) denotes the recruitment
deviation in year \(t\) with a variance of \(\sigma^{2}_{R}\).
\(Z_{a,t}\) represents the total mortality rate for fish of age \(a\) in
year \(t\), which is the summation of the fishing mortality \(F_{a,t}\)
and the constant natural mortality \(M\) (i.e., \(Z_{a,t}=F_{a,t}+M\)).
Finally, \(A\) represents the terminal age class, referred to as ``the
plus group,'' where fish that are the same age or older are aggregated.
\citet{jung2021study} showed that ages from 0 to 5 constituted the
majority of the catch in 2015 from their age-length key analysis, where
a small number of otolith samples (7 out of 401) were estimated to be
age of 6 and the growth increment between ages 5 and 6 was negligible.
Therefore, we assumed the terminal age class to be 5. The natural
mortality rate \(M\) was set to a fixed value of 0.53, which corresponds
to the median estimate of the natural mortality rate for chub mackerel
based on a review of various studies conducted by the Technical Working
Group of the North Pacific Fisheries Commission
\citep{nishijima2021update}. The initial abundance of age \(a\) (i.e.,
year \(t=1950\)) was assumed to be at near unfished equilibrium (denoted
as \(N_{a,0}\)), given that the catch for the initial two decades was
minimal compared to the later years of the time series (Fig.
\ref{MackData}a). The equation describing this initial abundance is
provided in Appendix A.

According to the separability assumption \citep{doubleday1976}, the
age-dependent fishing mortality \(F_{a,t}\) was calculated as the
product of the fully selected fishing mortality rate \(F_{t}\) (i.e.,
\(F_{t}=\max(F_{a,t})\)) and the age-dependent selectivity \(S_{a,t}\):

\begin{equation*}
\begin{aligned}
F_{a,t}=F_{t} \cdot S_{a,t}.
\end{aligned} 
\end{equation*}

The fully selected fishing mortality rate \(F_{t}\) was modeled using a
random walk process in log space \citep{nielsen2014estimation}:

\begin{equation*}
\begin{aligned}
\log(F_{t}) =
\begin{cases}
\log(F_{0}) + \varepsilon^{F}_{t}, \quad & \text{for } t=1950 \\ 
\log(F_{t-1}) + \varepsilon^{F}_{t}, \quad & \text{for } 1950 < t \leq 2021 
\end{cases},
\end{aligned} 
\end{equation*}

where \(\varepsilon^{F}_{t}\) represents the inter-annual deviation in
the fully selected fishing mortality rate in year \(t\), which was
assumed to follow a normal distribution with a mean of 0 and a variance
of \(\sigma^{2}_{F}\). We made the assumption that the stock was near
its unfished status in the initial year. Therefore, the fully selected
fishing mortality rate in the initial year, denoted as \(F_{0}\), was
considered negligible and fixed at 0.01.

Given the abundance of age \(a\) in year \(t\), the annual SSB
(\(\text{SSB}_{t}\)) was calculated as

\begin{equation*}
\text{SSB}_{t}= \varphi \cdot \sum_{a} N_{a,t} \cdot e^{-\phi \cdot Z_{a,t}} \cdot W_{a} \cdot m_{a}, 
\end{equation*}

where \(\varphi\) is the female proportion, \(W_{a}\) is the mean
weight-at-age, \(m_{a}\) is the maturity-at-age (proportion of mature
individuals at age \(a\)), and \(\phi\) is the fraction of the year
elapsed at the time of spawning. Detailed information regarding the
models and input values for \(W_{a}\), \(m_{a}\), and other parameter
inputs is provided in Table \ref{table:submodels}.

\subsubsection{Selectivity}\label{selectivity}

Several factors can contribute to variations in fisheries selectivity
over time, and neglecting these changes in stock assessment models can
result in biased estimates of abundance and mortality rates
\citep{gudmundsson2012selection}. Previous studies have highlighted that
assuming constant selectivity over time, when it is actually
time-varying, can lead to retrospective patterns in time series
quantities such as SSB and fishing mortality
\citep{martell2014towards, hurtado2015looking, szuwalski2018reducing}.

As discussed by previous studies
\citep{martell2014towards, hoyle2024catch}, we considered that
selectivity of the LPS fishery was time-varying and consisted of the two
main components: (i) availability of fish to the fishing gear,
influenced by factors such as spatial distribution of fish and
variations in fishing location and effort, and (ii) retention
probability of the fishing gear (i.e., the probability of capturing a
fish of a given size given that the fish is available to the fishing
gear).

We assumed that a time-varying component of selectivity in the LPS
fishery, as indicated by the temporal changes in the left limb (i.e.,
small size groups) of the length composition data (see Fig.
\ref{MackData}c), was primarily driven by changes in the availability of
young fish to the fishing gear. However, we made the assumption that the
retention probability of the fishing gear remained constant over time.
This allowed us to model time-invariant length-at-age probability
distributions in catch (see Eq. \eqref{eq:LengAtAgeDistCatch} in the
\textit{Length-at-age distributions in catch} section below for more
details). This time-invariant assumption for length-at-age distributions
was commonly made in other similarly structured models
\citep{fournier1998multifan, francis2016growth}.

To incorporate the time-varying component of selectivity in the LPS
fishery, we developed a model for age-dependent, time-varying
selectivity, denoted \(S_{a,t}\):

\begin{equation} 
\begin{aligned}
S_{a,t}=\dfrac{19}{19+e^{-\log(361) \cdot \left( \frac{a-a_{95}}{a_{95}-a_{05,t}} \right) }}.
\end{aligned} \label{eq:SelFunction}
\end{equation}

In this model, \(a_{05,t}\) represents the age at which 5\% selectivity
occurs in year \(t\). This value was assumed to either remain constant
(i.e., the same for all years) or vary with time, depending on chosen
model configurations (See the \textit{Alternative model configurations}
section below for more details). In contrast, \(a_{95}\) denotes the age
at which 95\% selectivity occurs, and this value was considered constant
over time. Such a time-invariant assumption was made, assuming that
selectivity for older age groups was relatively stable over time. This
is due to the common preference in commercial fisheries for targeting
large individuals. In sensitivity analyses where the constant age at 5\%
selectivity was assumed, we replaced \(a_{05,t}\) with the median age at
5\% selectivity, denoted as \(a_{05}\).

The time-varying age at 5\% selectivity was modeled using a bounded
logit-normal distribution:

\begin{equation}
\begin{aligned}
a_{05,t}=l_{a_{05}}+(u_{a_{05}}-l_{a_{05}}) \cdot \mathrm{logit}^{-1}(x_{t});  \quad x_{t}=  \mathrm{logit} \left( \dfrac{a_{05} - l_{a_{05}} }{u_{a_{05}}-l_{a_{05}} } \right) + \varepsilon^{S}_{t},
\end{aligned} \label{eq:a05t}
\end{equation}

where \(u_{a_{05}}\) and \(l_{a_{05}}\) denote the upper and lower
bounds for \(a_{05,t}\), respectively, \(\varepsilon^{S}_{t}\)
represents the deviation between \(a_{05,t}\) and \(a_{05}\). The vector
of the deviations \(\boldsymbol{\varepsilon}^{S}\) was assumed to follow
a multivariate normal distribution with a mean vector \(\boldsymbol{0}\)
and a covariance matrix \(\boldsymbol{\varSigma}\), where the covariance
matrix was modeled for a stationary AR1 process with a correlation
coefficient for consecutive deviations \(\varrho\):

\begin{equation}
\begin{aligned}
\boldsymbol{\varepsilon}^{S} \sim \text{MVN}(\boldsymbol{0}, \boldsymbol{\varSigma}); \quad \boldsymbol{\varSigma}_{t, \tilde{t}}=\dfrac{\sigma^{2}_{S}}{1-\varrho^{2}} \cdot \varrho^{|t-\tilde{t}|},
\end{aligned} \label{eq:a05Deviation}
\end{equation}

where \(\sigma^{2}_{S}\) is the variance of the deviation for each year
\(t\). The auto-correlation coefficient \(\varrho\) was not estimated,
but in sensitivity analyses was assumed to be one of 0, 0.3, 0.6, or
0.9.

\subsubsection{Catchability}\label{catchability}

To investigate the performance of the assessment model under diverse
catchability assumptions, we modeled catchability to reflect three
distinct scenarios: constant, linearly increasing, and undergoing a
random walk process in log space. The ``linear increase'' scenario was
chosen to simulate the impact of technological advancements in fishing
practices, such as improvements in vessel power and gear technology, as
documented by \cite{seo2017change}. This trend aims to capture the
gradual increase in fishing efficiency over time. On the other hand, the
``random walk'' scenario was designed to encapsulate the potential for
random fluctuations in catchability that could arise from factors not
directly related to technological progress, thus providing a more
comprehensive representation of variability in catchability. The
catchability was modeled as follows:

\begin{equation}
\begin{aligned}
\begin{cases}
q_{t}=q_{0}, \qquad  \text{for } 1976 \leq t \leq 2019 , & \text{ if constant } \\
q_{t}=
\begin{cases}
q_{0}, & \text{ for } t=1976 \\
q_{t-1}+c, & \text{ for } 1976 < t \leq 2019 \\
\end{cases},  & \text{ if linear increase} \\ 
\log(q_{t})=
\begin{cases}
\log(q_{0}), & \text{ for } t=1976  \\
\log(q_{t-1}) + \varepsilon^{q}_{t},    & \text{ for } 1976 < t \leq 2019  \\
\end{cases}, & \text{ if random walk}
\\ 
\end{cases},
\end{aligned} \label{eq:qModels} 
\end{equation}

where \(q_{0}\) is either the time-invariant catchability for the
``constant'' assumption or the initial catchability for the ``linear
increase'' and ``random walk'' assumptions, \(c\) is an additive fixed
annual increase, and \(\varepsilon^{q}_{t}\) is the inter-annual
deviation, which was assumed to follow a normal distribution with mean 0
and variance \(\sigma^{2}_{q}\).

\subsubsection{Length-at-age distributions in
catch}\label{length-at-age-distributions-in-catch}

An age-length key table, based on samples collected in 2015 from the LPS
fishery, is publicly available, as presented in \citet{jung2021study}.
\citet{gim2020management} utilized this information to convert length
frequencies into age frequencies for all observed years of length data.
They then applied an age-structured model to estimate stock status and
management quantities. It is crucial to note, however, that this
approach may introduce significant bias into estimates of stock size and
recruitment. This potential bias arises because the age-length key table
encodes the proportion of age groups within each length bin, which
should be year-specific, mainly due to recruitment variability
\citep{kimura1977statistical, westrheim1978bias, ailloud2019general}.
This seemingly straightforward application overlooks the fundamental
difference between the length distribution given age (\(\pi_{j|a}\)) and
the age distribution given length (\(\pi_{a|j}\))
\citep{ailloud2019general}.

To address this issue, we derived the length distribution given age
(\(\pi_{j|a}\)) using the available information on the minimum, maximum,
and mean lengths of fish for each age group, as provided by
\citet{jung2021study} (see Table \ref{table:submodels} for details).
Since this information is derived from fishery catch data rather than
directly from the population, we made an assumption that the length
distribution follows a skew normal distribution
\citep{azzalini1985class}, rather than a normal distribution, in order
to account for selection effects in the fishery catch.

We matched the minimum, maximum, and mean lengths of fish for each age
group with the 5th, 95th percentiles, and the mean of a skew normal
distribution. Utilizing these matches, we derived the parameters of a
skew normal distribution and applied the estimated parameters to compute
the length-at-age distributions in catch, serving as input for the
assessment model. The length-at-age probability distribution in catch
(i.e., the probability of a fish of age \(a\) being in the \(j\)th
length bin, given that it was retained in catch), denoted as
\(\pi_{j|a}\), was calculated as follows:

\begin{equation}
\begin{aligned}
\pi_{j|a} =
\begin{cases}
\Phi(L_{j}+0.5 \cdot d |\xi_{a}, \omega_{a}, \alpha_{a}), & \text{for } j=1 \\
\Phi(L_{j}+0.5 \cdot d |\xi_{a}, \omega_{a}, \alpha_{a})-\Phi(L_{j}-0.5 \cdot d|\xi_{a}, \omega_{a}, \alpha_{a}), & \text{for } 1<j < J \\
1-\Phi(L_{j}-0.5 \cdot d|\xi_{a}, \omega_{a}, \alpha_{a}), & \text{for } j=J
\end{cases},
\end{aligned} \label{eq:LengAtAgeDistCatch}
\end{equation}

where \(\Phi (x|\cdot)\) is the cumulative distribution function (CDF)
of the skew normal distribution for length \(L\) less than or equal to
\(x\) (see Appendix B for the definition of its CDF), \(L_{j}\) is the
midpoint of the length bin \(j\), \(d\) is the length bin width (we set
\(d=1\)), and \(\xi_{a}\), \(\omega_{a}\), and \(\alpha_{a}\) are the
location, scale, and shape parameters of the skew normal distribution
for length-at-age \(a\), respectively.

\subsection{Observation models}\label{observation-models}

The observation model components illustrate how the observed CPUE values
(\(I_{t}\)), annual catch values (\(Y_{t}\)), and length composition
data (\(\boldsymbol{n}_{t}\)) are linked to the age-structured
population dynamics. Detailed descriptions of each observation model are
provided in the following sections.

\subsubsection{CPUE}\label{cpue}

The natural logarithm of the CPUE \(\log(I_{t})\) was assumed to follow
a normal distribution with mean \(\log(\widehat{I}_{t})\) and variance
\(\sigma^{2}_{I}\):

\begin{equation*}
\begin{aligned}
\log(I_{t}) \sim \text{Normal} (\log(\widehat{I}_{t}), \sigma^{2}_{I}),
\end{aligned} 
\end{equation*}

where \(\widehat{I}_{t}\) is the model-predicted CPUE in year \(t\):

\begin{equation*}
\begin{aligned}
\widehat{I}_{t}=q_{t} \cdot \text{VB}_{t},
\end{aligned} 
\end{equation*}

in which \(q_{t}\) is the catchability coefficient in year \(t\), and
\(\text{VB}_{t}\) is the biomass vulnerable to the fishery in year
\(t\). The vulnerable biomass \(\text{VB}_{t}\) was calculated as
follows:

\begin{equation*}
\begin{aligned}
\text{VB}_{t}=\sum^{A}_{a=1} N_{a,t} \cdot e^{-0.5 \cdot Z_{a,t}}  \cdot S_{a,t} \cdot W_{a}.
\end{aligned} 
\end{equation*}

\subsubsection{Annual catch}\label{annual-catch}

The natural logarithm of the annual catch (in weight) \(\log(Y_{t})\)
was assumed to follow a normal distribution with mean
\(\log(\widehat{Y}_{t})\) and variance \(\sigma^{2}_{Y}\):

\begin{equation*}
\begin{aligned}
\log(Y_{t}) \sim \text{Normal} (\log(\widehat{Y}_{t}),  \sigma^{2}_{Y}),
\end{aligned} 
\end{equation*}

where \(\widehat{Y}_{t}\) is the model-predicted annual catch in year
\(t\):

\begin{equation*}
\begin{aligned}
\widehat{Y}_{t} = \sum^{A}_{a=1} \widehat{C}_{a,t} \cdot W_{a}.
\end{aligned} 
\end{equation*}

The model-predicted age-specific catch in numbers \(\widehat{C}_{a,t}\)
was calculated using the Baranov catch equation
\citep{baranov1918question}:

\begin{equation}
\begin{aligned}
\widehat{C}_{a,t} = \dfrac{F_{a,t}}{Z_{a,t}} \cdot N_{a,t} \cdot (1-e^{-Z_{a,t}}).
\end{aligned} \label{eq:CatchAtAge}
\end{equation}

\subsubsection{Length composition}\label{length-composition}

A simulation study by \citet{xu2020comparing} demonstrated that a
Dirichlet-multinomial (DM) distribution outperforms other commonly used
data weighting methods for composition data, when time-varying
selectivity is present. Thus, we employed a DM distribution to model the
length composition data:

\begin{equation*}
\boldsymbol{n}_{t} \sim \text{DM}(E_{t}, \boldsymbol{\delta}_{t}), \quad \text{for } 2000 \leq t \leq 2017,
\end{equation*}

where \(\boldsymbol{n}_{t}\) represents the vector of observed fish
counts collected for each length bin \(j\) in year \(t\), \(E_{t}\)
denotes the sample size of the length composition observations in year
\(t\), and \(\boldsymbol{\delta}_{t}\) is the vector of concentration
parameters for year \(t\) (defined as how concentrated the observed
length composition proportions are around the model-predicted length
composition proportions).

Following \citet{thorson2017model}, we parameterized the concentration
parameter for each length bin \(j\) in year \(t\) (i.e.,
\(\delta_{j,t}\)) as a function of the corresponding model-predicted
length composition proportion \(\widehat{P}_{j|t}\) and the
variance-inflation parameter \(\beta_{t}\):

\begin{equation*} 
\delta_{j,t}= \beta_{t} \cdot \widehat{P}_{j|t},
\end{equation*}

where the variance-inflation parameter \(\beta_{t}\) was assumed to have
a linear relationship with both the sample size \(E_{t}\) and the scale
parameter \(\theta\) (i.e., \(\beta_{t}=E_{t} \cdot \theta\)). This
assumption led to the formulation of the effective sample size as
\(E^{\text{eff}}_{t}=(1+\theta \cdot E_{t})/(1+\theta)\), where
\(E^{\text{eff}}_{t}\) represents the effective sample size for
observations in year \(t\) (for further details, refer to
\citet{thorson2017model}).

The model-predicted length composition proportion \(\widehat{P}_{j|t}\)
was derived as follows:

\begin{equation*}
\begin{aligned}
\widehat{P}_{j|t} = \dfrac{\widehat{C}_{j,t}}{\sum^{J}_{j'=1} \widehat{C}_{j',t}},
\end{aligned} 
\end{equation*}

where \(\widehat{C}_{j,t}\) represents the model-predicted catch (in
numbers) of fish in length bin \(j\) for year \(t\). This catch was
derived by converting the age-specific catch \(\widehat{C}_{a,t}\) in
Eq. \eqref{eq:CatchAtAge} to the length-specific catch
\(\widehat{C}_{j,t}\), based on the length-at-age distribution in catch
(i.e., \(\pi_{j|a}\) in Eq. \eqref{eq:LengAtAgeDistCatch}):

\begin{equation*}
\begin{aligned}
\widehat{C}_{j,t} = \sum^{A}_{a=1} \widehat{C}_{a,t} \cdot \pi_{j|a}.
\end{aligned} 
\end{equation*}

\subsection{Alternative model
configurations}\label{alternative-model-configurations}

To explore the effects of different model configurations on the model
performance, we fitted a total of 15 models (denoted as M1-15) to the
data, each differing in several key aspects:

\begin{itemize}
      \item The selectivity was either constant (i.e., setting $a_{05,t}=a_{05}$ for all $t$ in Eq. \eqref{eq:SelFunction}) or time-varying (i.e., allowing $a_{05,t}$ in Eq \eqref{eq:a05t} to vary over time).
      \item The auto-correlation coefficient (i.e., $\varrho$ in Eq. \eqref{eq:a05Deviation}) for deviations of age at 5\% selectivity was fixed at 0.0, 0.3, 0.6, or 0.9 (note that we initially attempted to estimate $\varrho$ as part of the model fitting process, but it was found to be non-identifiable and was thus fixed at these values).
      \item The catchability coefficient was either constant or time-varying in two distinct ways (i.e., linearly increasing or following a random walk), as specified in Eq. \eqref{eq:qModels}.
\end{itemize}

Details of the model configurations are summarized in Table
\ref{table:models_config}.

\subsection{Standard deviation penalty
function}\label{standard-deviation-penalty-function}

Estimating both process and observation standard deviation (SD)
parameters in state-space models is a challenging task, as highlighted
in previous works
\citep{auger2016state, auger2021guide, hyun2022evaluation, kim2022state}.
Even in a simple random walk model, these SD parameters are practically
non-identifiable without additional constraints \citep{auger2016state}.
Non-identifiability of those parameters is often indicated by boundary
estimates
\citep{auger2016state, auger2021guide, hyun2022evaluation, kim2022state}
and a flat likelihood surface near the boundary
\citep{raue2009structural, auger2016state, auger2021guide}. This
non-identifiability can lead to biased estimates of the other parameters
and infinite confidence intervals
\citep{raue2009structural, auger2016state, auger2021guide, hyun2022evaluation, kim2022state}.

In our study, we aimed to simultaneously estimate both process and
observation variance parameters. To accomplish this, we adopted the
methodology introduced by \citet{chung2013nondegenerate}, applying gamma
distribution penalties to the SD parameters associated with the
observation models (i.e., \(\sigma_{Y}\), \(\sigma_{I}\), and
\(\sigma_{q}\)).

The gamma distribution, serving as a penalty function for
\(\sigma_{X\in {\{Y, I, q\} }}\), was defined as follows:

\begin{equation*}
\begin{aligned}
f(\sigma_{X})=\frac{\lambda^{\psi}}{\Gamma(\psi)} \cdot \sigma_{X}^{\psi-1} \cdot e^{-\lambda \cdot \sigma_{X}},
\end{aligned}
\end{equation*}

where \(\psi\) is the shape parameter and \(\lambda\) is the rate
parameter. \citet{chung2013nondegenerate} recommended setting \(\psi=2\)
and \(\lambda \rightarrow 0\) (to achieve this, we used a extremely
small value for \(\lambda\), i.e., \(\lambda=10^{-10}\)). This choice of
\(\psi=2\) ensures positive estimates in maximum likelihood estimation
for \(\sigma_{X}\), as the gamma distribution is zero at the origin.
Furthermore, as \(\lambda\) approaches zero, the gamma distribution
function exhibits a positive constant derivative at zero, allowing the
likelihood to dominate even in the presence of strong curvature near
zero \citep{chung2013nondegenerate}.

The joint likelihoods of the models incorporating these gamma penalties
are outlined in Table \ref{table:models_likelihood}. We conducted a
comparison between the models with and without the gamma penalties to
evaluate the efficacy of this penalized estimation method.

\subsection{Model estimation}\label{model-estimation}

The estimation process was carried out using Template Model Builder
(\texttt{TMB}) \citep{kristensen2015tmb} in \texttt{R}
\citep{Rcore2023}. The model parameters \(\boldsymbol{\Theta}\) were
estimated by maximizing the marginal likelihood of the data
\(\mathcal{L} (\boldsymbol{\Theta} | \boldsymbol{D})\) using the
\texttt{nlminb} optimizer in \texttt{R}.

In \texttt{TMB}, the marginal likelihood of the data
\(\mathcal{L} (\boldsymbol{\Theta} | \boldsymbol{D})\) was obtained by
integrating out the random effects \(\boldsymbol{Q}\) from the joint
likelihood
\(\mathcal{L} (\boldsymbol{\Theta}, \boldsymbol{Q} |\boldsymbol{D})\)
using the Laplace approximation technique
\citep{skaug2006automatic, kristensen2015tmb}:

\begin{equation*}
\begin{aligned} 
\mathcal{L} (\boldsymbol{\Theta} | \boldsymbol{D})  = \int \mathcal{L} (\boldsymbol{\Theta}, \boldsymbol{Q} |\boldsymbol{D}) d\boldsymbol{Q}.
\end{aligned}
\end{equation*}

Through the maximization of the marginal likelihood concerning the model
parameters \(\boldsymbol{\Theta}\), we obtained the estimates denoted as
\(\widehat{\boldsymbol{\Theta}}\):

\begin{equation*}
\begin{aligned} 
\widehat{\boldsymbol{\Theta}}   = \text{arg}~\underset{\boldsymbol{\Theta}}{\text{max}}~\log[\mathcal{L} (\boldsymbol{\Theta} |\boldsymbol{D})].
\end{aligned}
\end{equation*}

Once \(\widehat{\boldsymbol{\Theta}}\) was determined, \texttt{TMB}
proceeded to estimate the random effects \(\boldsymbol{Q}\) by
maximizing the joint likelihood with respect to \(\boldsymbol{Q}\),
while keeping \(\boldsymbol{\Theta}\) fixed at
\(\widehat{\boldsymbol{\Theta}}\):

\begin{equation*}
\begin{aligned} 
\widehat{\boldsymbol{Q}}   = \text{arg}~\underset{\boldsymbol{Q}}{\text{max}}~\log[\mathcal{L} (\widehat{\boldsymbol{\Theta}}, \boldsymbol{Q} |\boldsymbol{D})].
\end{aligned}
\end{equation*}

The uncertainties of the parameter estimates were assessed using the
delta method, which involved calculating the determinant of the Hessian
matrix of the marginal likelihood at \(\widehat{\boldsymbol{\Theta}}\)
through numerical Cholesky decomposition
\citep{skaug2006automatic, kristensen2015tmb}. To determine successful
model convergence, we checked if the absolute value of the maximum
gradient component of the parameters was close to 0 (i.e., less than
0.1), and that the Hessian matrix was positive definite. The joint
likelihoods for all models considered in this study are defined in Table
\ref{table:models_likelihood}.

\subsection{Model performance
evaluation}\label{model-performance-evaluation}

The relative performance of all alternative models was assessed using
Akaike's Information Criterion (AIC) \citep{akaike1974new}. AIC was
calculated using the maximized marginal log-likelihoods of the models,
where random effects were integrated out but the gamma distribution
penalties were included, and the number of fixed-effect parameters
(i.e., \(\text{AIC} = -2 \cdot \log( \mathcal{L}_{MLE} ) +2 \cdot k\),
where \(\mathcal{L}_{MLE}\) represents the maximized marginal
log-likelihood and \(k\) is the number of fixed-effect parameters).

Previous studies have demonstrated that the marginal likelihood-based
AIC is reliable for selecting the best-performing model in state-space
models \citep{auger2017spatiotemporal, miller2018evaluating}. To further
validate the accuracy of AIC in model selection, we conducted cross-test
simulations. In these simulations, we treated each candidate model as an
operating model (OM) to generate pseudo data. Subsequently, all
candidate models, now serving as estimation models (EMs), were fitted to
this pseudo data. We then compared the AIC values of these models to
ascertain whether the matching OM and EM combination yielded the lowest
AIC value. Because of the large number of candidate models in this
factorial experiment, a subset of alternative models (namely M10, M11,
and M12) was selected for the cross-test simulations.

\subsubsection{Robustness of the model to observation
error}\label{robustness-of-the-model-to-observation-error}

\citet{deroba2015simulation} highlighted the common occurrence of
divergence in self-tests for stock assessment models employing random
walks for time-varying processes, particularly in the most recent years
of the time series. They recommended conducting a self-consistency test
to ensure that the model was correctly specified. In line with this
recommendation, we conducted a simulation test to assess the model's
robustness to observation errors.

In this simulation, we generated a pseudo dataset using the model, with
parameters and random effects fixed at the estimates obtained from
fitting the mackerel data. Observation errors and length frequencies
were drawn randomly from the assumed distributions. Subsequently, we
fitted the simulated dataset to the same model to re-estimate the
parameters and random effects.

To ensure the reliability of the model, we performed 500
simulation-estimation runs for this self-consistency check. The model
was considered robust if the median of the estimates from the
simulation-estimation runs closely matched the true values used for
simulating the pseudo dataset \citep{deroba2015simulation}. We visually
assessed the divergence between the true and estimated values by
comparing summary statistics (e.g., 2.5\%, 50\%, and 97.5\% percentiles)
of the estimated time series of key quantities of interest (e.g., total
biomass, SSB, fishing mortality, and catchability) with those of the
true values.

\subsubsection{Retrospective analysis}\label{retrospective-analysis}

We examined retrospective patterns to identify potential model
misspecification. In integrated stock assessment models,
misspecification is typically indicated by systematic changes in
estimates of time series quantities when the model is fitted to data,
with the sequential exclusion of the most recent data points
\citep{hurtado2015looking}.

We performed a retrospective analysis by sequentially excluding the last
8 years of the data and fitting the model to the reduced dataset. To
visually assess retrospective patterns, we calculated the relative
difference (RD) between the estimates of time series quantities obtained
from the model using the full dataset and those from the models using
the reduced datasets. The RD for year \(t\) of the time series quantity
\(\zeta\) after removing \(y\) years of data (i.e.,
\(\text{RD}_{t,y}(\zeta)\)) was defined as follows:

\begin{equation*}
\begin{aligned}
\text{RD}_{t,y}(\zeta) = \dfrac{\widehat{\zeta}_{t,T-y }}{\widehat{\zeta}_{t,T}}-1, \quad \text{for } 1950 \leq t \leq T-y,
\end{aligned}
\end{equation*}

where \(T\) represents the terminal year of the catch data, \(y\)
indicates the number of years removed from the terminal year, which
spans from 1 to \(g\) (where \(g=8\) in this study),
\(\widehat{\zeta}_{t,T-y}\) denotes the estimate of the quantity
\(\zeta\) for year \(t\) derived from the model fitted to data from 1950
to \(T-y\), while \(\widehat{\zeta}_{t,T}\) represents the corresponding
estimate from the model fitted to the full dataset. In our retrospective
analysis, we considered \(\text{SSB}_{t}\) and \(F_{t}\) as the
quantities \(\zeta\).

In addition to visual inspection, we quantified the retrospective
pattern using Mohn's \(\rho\) \citep{mohn1999retrospective} for the time
series quantity \(\zeta\) (i.e., \(\rho(\zeta)\)). Mohn's \(\rho\) was
defined as follows:

\begin{equation}
\begin{aligned}
\rho(\zeta) = \dfrac{1}{g} \sum^{g}_{y=1} \dfrac{\widehat{\zeta}_{T-y,T-y}}{\widehat{\zeta}_{T-y,T}}-1,
\end{aligned} \label{eq:mohn_rho}
\end{equation}

where \(\widehat{\zeta}_{T-y,T-y}\) represents the estimate of the
quantity \(\zeta\) for year \(T-y\) from the model fitted to the data
from 1950 to \(T-y\), \(\widehat{\zeta}_{T-y,T}\) is the corresponding
estimate from the model fitted to the full dataset. Given the
inconsistent time series lengths for catch data (1950 to 2021), CPUE
data (1976 to 2019), and length frequency data (2000 to 2017), we also
computed Mohn's \(\rho\) for the time period where all data types were
available. In this case, \(y\) ranged from 4 to \(g\), and the division
by \(g\) was replaced by \(g-3\).

Establishing a definitive threshold value for Mohn's \(\rho\) to detect
retrospective patterns remains a challenge in stock assessments.
Therefore, we followed the recommendations of
\citet{hurtado2015looking}, who proposed a practical rule of thumb for
species with shorter lifespans. They suggest that Mohn's \(\rho\) values
exceeding \(0.3\) or falling below \(-0.22\) can be indicative of
potential retrospective patterns deserving attention.

\subsubsection{Parametric bootstrap}\label{parametric-bootstrap}

To assess the bias, estimability, and uncertainty of the model
parameters under each model configuration, we conducted a parametric
bootstrap analysis, similar to the self-consistency test described
above. However, in this analysis, the pseudo-dataset was generated from
the model where the random effects were not fixed but drawn from the
assumed distributions with the estimated variance parameters.
Subsequently, the pseudo-dataset was fitted to the same model to
re-estimate the parameters and random effects, which were then compared
to the corresponding true values.

We conducted a total of 1000 simulation-estimation runs for this
parametric bootstrap analysis. In each run, we checked whether the 95\%
confidence intervals of the estimates included the true values to
investigate the 95\% coverage probability of the model. The Wald method
was used to compute the 95\% confidence intervals of the parameters.

The bias of the parameter estimates was quantified as the median of the
RD between the estimated and true values. The RD of a parameter
\(\Theta\) for the \(i\)th simulation-estimation run was calculated as
the ratio of the estimate \(\widehat{\Theta}\) to the true value
\(\Theta_{\text{true}}\), subtracting one:

\begin{equation}
\begin{aligned}
\text{RD}_{i}(\Theta) = \dfrac{\widehat{\Theta}_{i}}{\Theta_{\text{true}}}-1
\end{aligned} \label{eq:RDpars}
\end{equation}

We also assessed the estimability of the parameters by analyzing the
distribution of parameter estimates obtained from the
simulation-estimation runs. Inestimable parameters often exhibit
multi-modal distributions, and variance parameters, especially those in
state-space models, frequently display a bimodal distribution, one mode
of which is near zero, as highlighted by previous studies
\citep{auger2016state, auger2021guide, hyun2022evaluation, kim2022state}.

\subsubsection{Random effects
validation}\label{random-effects-validation}

We evaluated the distributional assumptions of the random effects
\(\boldsymbol{Q}\) using a simulation-based approach
\citep{waagepetersen2006simulation, thygesen2017validation}. To validate
these assumptions, we generated 1000 sets of samples from the joint
posterior distribution of \(\boldsymbol{Q}\) and conducted an
examination.

The joint posterior distribution of \(\boldsymbol{Q}\) was approximated
using a multivariate normal distribution. The mean vector represented
the estimates of the random effects, denoted as
\(\widehat{\boldsymbol{Q}}\), and the covariance matrix was derived from
the negative inverse of the Hessian matrix
\citep{thygesen2017validation}. The samples were then normalized using
the estimated SD of the random effects, such as
\(\widehat{\sigma}_{R}\), \(\widehat{\sigma}_{F}\),
\(\widehat{\sigma}_{S}\), and \(\widehat{\sigma}_{q}\). For
auto-correlated random effects in selectivity deviations (i.e.,
\(\boldsymbol{\varepsilon}^{S}\) in M7-15), we multiplied the inverse of
the Cholesky decomposition of the corresponding covariance matrix to
normalize and decorrelate the samples.

To validate the correct specification of our model, we investigated
whether the normalized samples conform to a standard normal
distribution. This evaluation was conducted visually through
quantile-quantile (Q-Q) plots. The challenge in interpreting these plots
lies in accurately distinguishing between deviations from expected
values caused by random sampling variation and actual violations of
assumptions. To explore systematic deviations from a standard normal
distribution, we analyzed 1000 sets of Q-Q plots. Our objective was to
establish a 95\% envelope for these plots, following
\citet{rubio2016bayesian}. We visually assessed whether the 95\%
envelope contained the straight line with an intercept of 0 and a slope
of 1, representing a perfect adherence to the normality assumption.

\section{Results}\label{results}

All 15 models (M1-15) successfully converged. The initial values and
parameter bounds used for each model are summarized in Supplementary
Table \ref{table:bounds}. The estimates of the parameters for the
length-at-age distributions in catch are given in Table
\ref{table:skew_normal_catch} (see Supplementary Fig. \ref{LengthAge}
for their visualized forms). The fitted estimates compared to the
observed data are presented in Supplementary Figs. \ref{FittedValues}
and \ref{FittedValuesLF}. The estimated time-varying age at 5\%
selectivity and recruitment deviations are shown in Supplementary Fig.
\ref{SelRecEstim}. Based on AIC and various model validation criteria
that we used in this study (details are provided in following sections),
the best-performing model was M11, characterized by time-varying
selectivity with an auto-correlation coefficient \(\varrho=0.6\) and a
linearly increasing catchability. Other models with identical
configurations, varying only in the \(\varrho\) values, also exhibited
lower AIC values when compared to the models with time-invariant
selectivity (M1-3) and those with time-varying selectivity but with
random walk catchability (M6, M9, M12, M15). This tendency in AIC values
suggested that model selection was less sensitive to the choice of
\(\varrho\) and more influenced by the choice of the catchability
function combined with time-varying selectivity. The differences in
terms of AIC between the best-performing model (M11) and other models
are summarized in Table \ref{table:models_config}.

\begin{figure}

{\centering \includegraphics[width=0.95\linewidth]{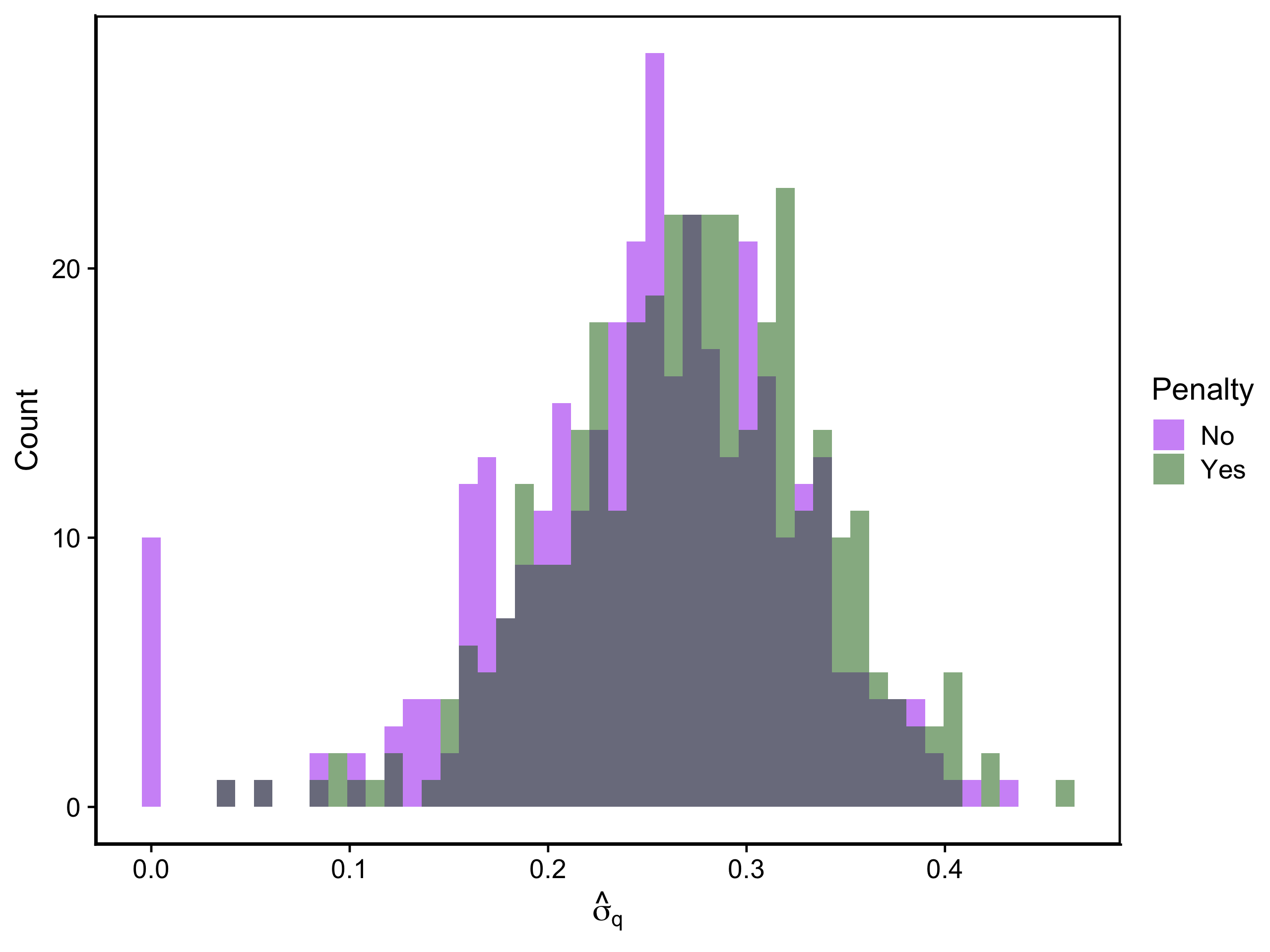} 

}

\caption{Histograms of the estimates of the standard deviation parameter for the random walk catchability ($\widehat{\sigma}_{q}$) in M12, with (dark green) and without (purple) the gamma penalty on the parameter. \label{MackSigq}}\label{fig:unnamed-chunk-39}
\end{figure}

The cross-test for checking the accuracy of AIC in model selection
revealed that AIC identified the matching model between the OM and the
EM with a high rate of success for M10 (400 out of 450) and M11 (466 out
of 467). However, when M12 was used as the OM, AIC failed to identify
the corresponding EM (only 1 out of 354 was successful). This poor
performance of AIC in model selection was attributed to the additional
gamma distribution penalty term in the likelihood function of M12,
imposed on \(\sigma_{q}\). When we removed the additional gamma
distribution penalty term from M12, AIC identified the matching model
with the EM even for M12 with a relatively high rate of success (306 out
of 355; see Supplementary Table \ref{table:cross-test} for the detailed
results). Despite the improvement, the distribution of the estimates of
\(\sigma_{q}\) showed a clear bimodal pattern (Supplementary Fig.
\ref{MackSigq}), indicating that the model was not well-identified. This
result suggested that the additional gamma distribution penalty term was
necessary for the model to be well-identified, but it also made the
model selection based on AIC less reliable.

The various assumptions regarding catchability had a pronounced impact
on the overall trends in fishing mortality. Models assuming constant
catchability (M1, M4, M7, M10, M13) exhibited nearly monotonic
decreasing trends since 1985 (see panels in the first column in Fig.
\ref{MackSelfF}). In contrast, models assuming linearly increasing
catchability (M2, M5, M8, M11, M14) demonstrated relatively stable
fishing mortality trends from 1984 until 2012, followed by abrupt
increases until 2015 and subsequent decreases thereafter (see panels in
the second column in Fig. \ref{MackSelfF}). Models incorporating random
walk catchability (M3, M6, M9, M12, M15) revealed a pattern akin to
those assuming linearly increasing catchability for the last decade but
with gradual increasing trends from 1980 to 2000 (see panels in the
third column in Fig. \ref{MackSelfF}).

The catchability trends under the random walk assumption (M3, M6, M9,
M12, M15) exhibited almost monotonic increases from 1990 to 2010,
followed by sudden increases from 2010 to 2015, and subsequent decreases
thereafter (see panels in the third column of Fig. \ref{MackSelfq}).
This pattern closely resembled the fishing mortality trends observed in
the same models.

\begin{figure}

{\centering \includegraphics[width=0.95\linewidth]{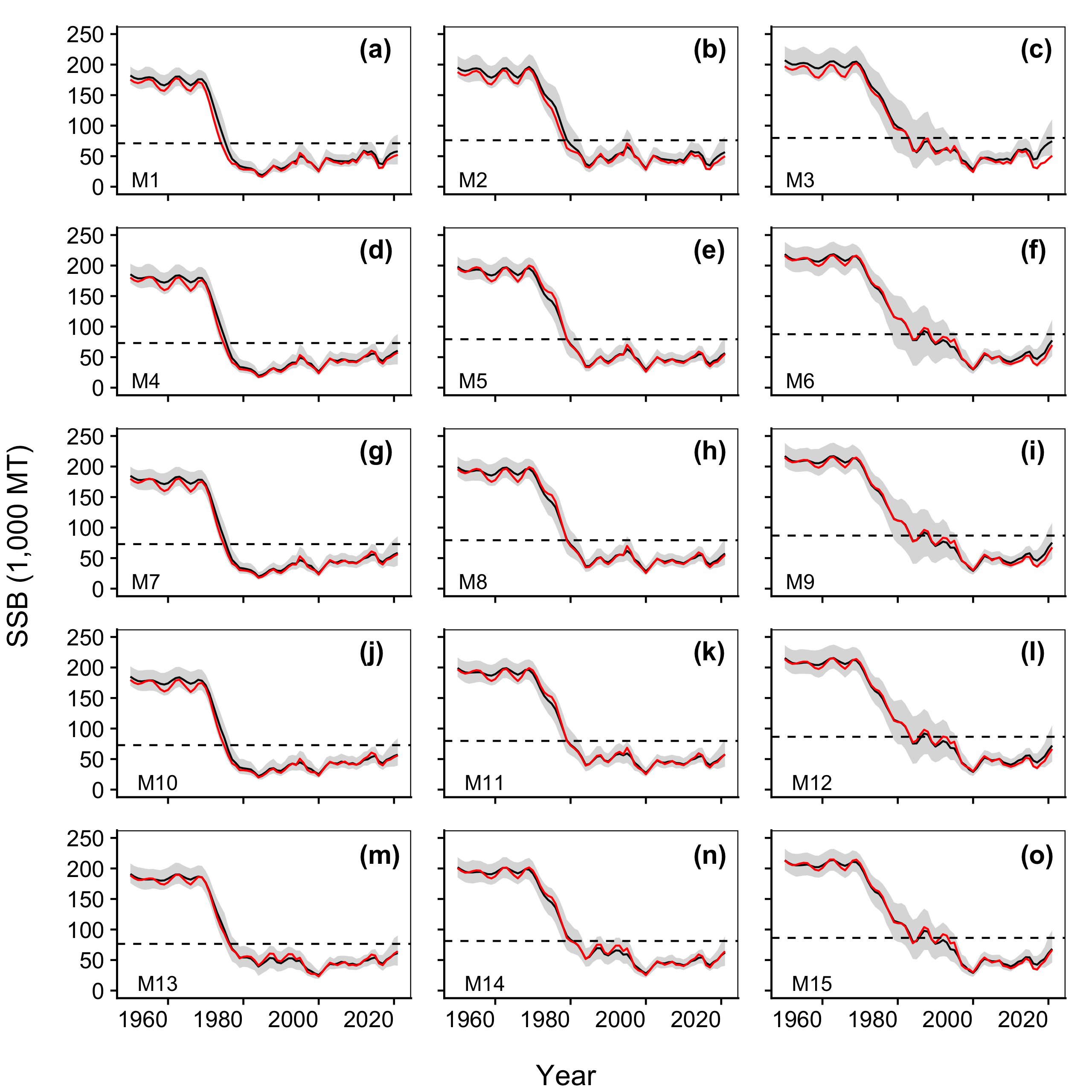} 

}

\caption{Self-test estimates of the spawning stock biomass (SSB) and corresponding true values for all alternative models (M1-15), with the model name indicated in the bottom-left corner of each panel. The black lines indicate the median values of the self-test estimates, and the gray areas represent the 95\% intervals. The red lines indicate the true values. The horizontal dashed lines denote the 40\% of the virgin spawning stock biomass level for each model. \label{MackSelfSSB}}\label{fig:unnamed-chunk-16}
\end{figure}

The SSB trends in all models exhibit a similar pattern. They were
characterized by stable fluctuations around an unfished status from 1950
to 1970, attributed to low catch levels during this period. This was
followed by a steep decrease until the early 1980s (or until 2000 in
models with random walk catchability), and subsequent stability around a
low SSB level (see Fig. \ref{MackSelfSSB}). Notably, models with
constant catchability showed a slight increase during this period of low
SSB, although the magnitude of the increase was not substantial (see
Fig. \ref{MackSelfSSB}). The SSB estimates from all models consistently
remained below 40\% of the unfished SSB (i.e.,
\(\text{SSB}_{t} < 0.4 \cdot \text{SSB}_{0}\), where \(\text{SSB}_{0}\)
represents the SSB at the unfished status and serves as a proxy for
interpreting the stock status in this study) for the last few decades
(see Fig. \ref{MackSelfSSB}). This indicated that, despite the
implementation of the TAC-based management system for this mackerel
stock since 1999 \citep{sim2020analysis}, the stock has not been managed
sustainably.

\subsection{Retrospective analysis}\label{retrospective-analysis-1}

\begin{figure}

{\centering \includegraphics[width=0.95\linewidth]{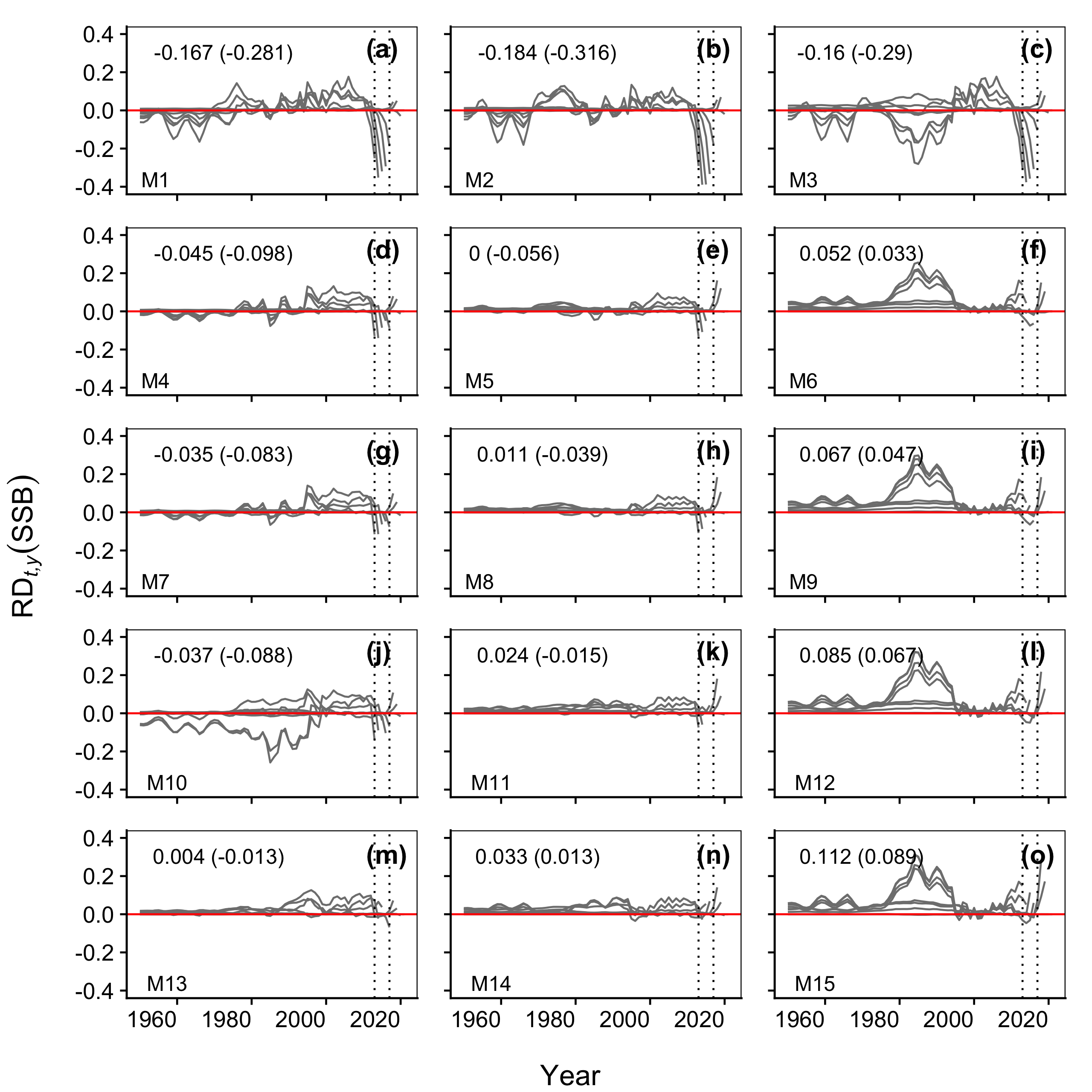} 

}

\caption{Relative difference (RD) measures (gray lines) for visualizing retrospective patterns in the estimated spawning stock biomass (SSB) for all alternative models (M1-15), with the model name indicated in the bottom-left corner of each panel. RD was calculated as the difference between estimates from the model fitted to the full dataset and the model fitted to the iteratively removed data for the last eight years, divided by the estimates from the model fitted to the full dataset. The horizontal red line indicates no difference between the two estimates, and the two vertical dotted lines mark the time period when all three datasets were available when fitting the model to the reduced dataset. The values at the top left corner of each panel indicate Mohn's $\rho$ values for all eight years of iterative data removal and fitting, with values in parentheses indicating the $\rho$ values for the five years of iterative data removal and fitting (2013 to 2017), a period during which complete data was available (between the two vertical dotted lines). \label{MackRetroSSB}}\label{fig:unnamed-chunk-17}
\end{figure}

\begin{figure}

{\centering \includegraphics[width=0.95\linewidth]{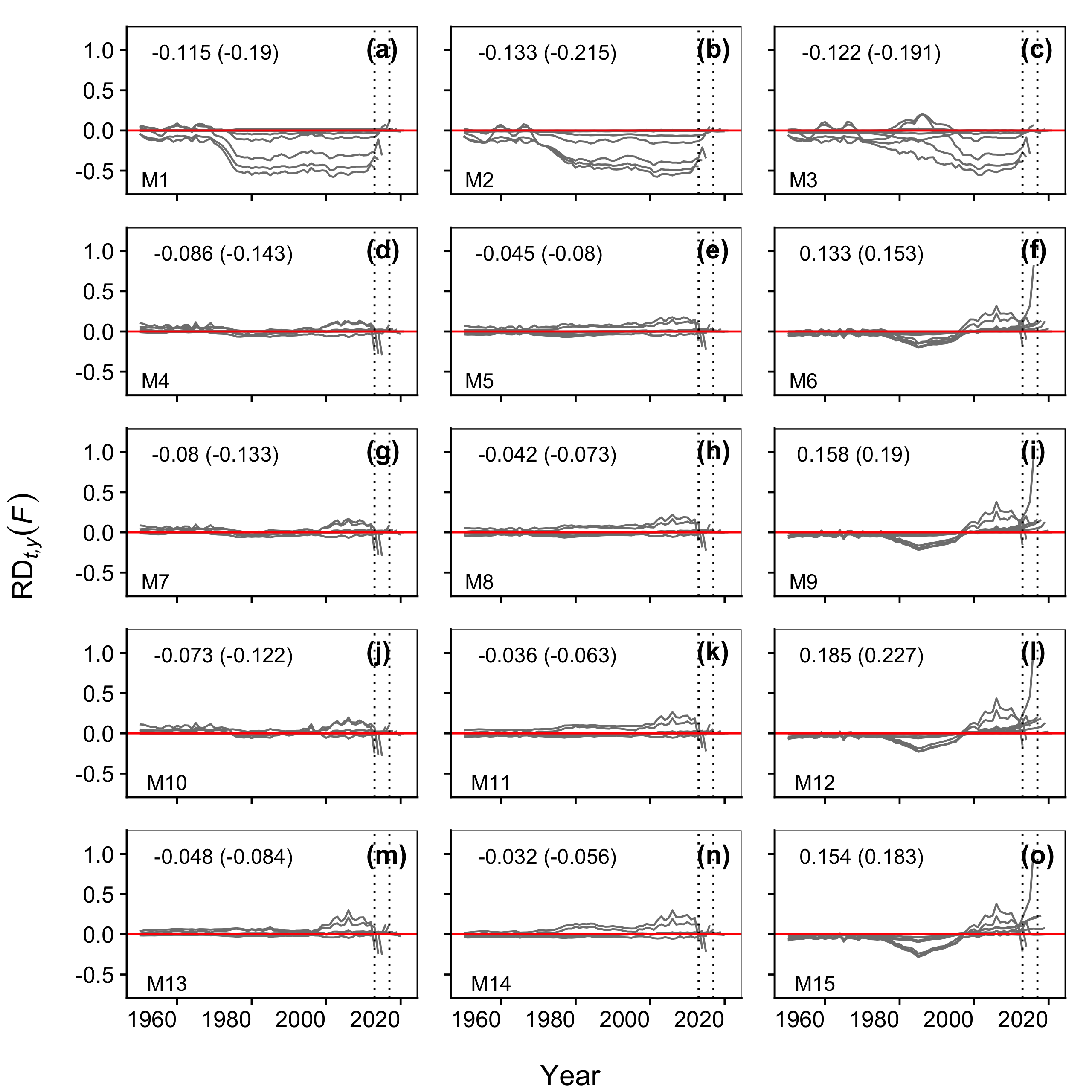} 

}

\caption{Relative difference (RD) measures (gray lines) for visualizing retrospective patterns in the estimated fishing mortality rate ($F$) for all alternative models (M1-15), with the model name indicated in the bottom-left corner of each panel. Other details about the calculation of RD are as described in Fig. \ref{MackRetroSSB}. \label{MackRetroF}}\label{fig:unnamed-chunk-18}
\end{figure}

Systematic retrospective patterns were observed in the models with
time-invariant selectivity (M1-3), as evidenced by the retrospective
discrepancy measures depicted in Figs. \ref{MackRetroSSB} and
\ref{MackRetroF}. In contrast, the models with time-varying selectivity
(M4-15) did not exhibit any notable retrospective patterns. This
observation suggested that the introduction of time-varying selectivity
effectively addressed model mis-specification.

Furthermore, the Mohn's \(\rho\) values derived from the SSB time-series
of those problematic models were all negative, specifically for M1-3
(i.e., \(\rho(\text{SSB})=-0.281,-0.316, -0.29\) for M1-3,
respectively). According to the guideline proposed by
\citet{hurtado2015looking}, where a Mohn's \(\rho\) value exceeding
\(0.3\) or falling below \(-0.22\) indicates the presence of
retrospective patterns, these values strongly suggested the existence of
significant retrospective patterns. The retrospective patterns were
particularly evident during the period when three different sets of data
were all available (see values in the parentheses in Figs.
\ref{MackRetroSSB} and \ref{MackRetroF}).

Although no systematic retrospective patterns were observed in the
models with time-varying selectivity and random walk catchability (M6,
M9, M12, M15) for the last eight years, there were indications of model
mis-specification. This was evident from the systematic dome-shaped
curve patterns in the SSB time-series (see Fig. \ref{MackRetroSSB}f, i,
l, and o) and the S-shaped curve patterns in the fishing mortality
time-series (see Fig. \ref{MackRetroF}f, i, l, and o). These systematic
patterns suggested that the models with random walk catchability were
influenced by certain data points when fitted to the reduced datasets.
This was further confirmed by the self-test results, where observation
errors were randomized to test the robustness of the models, revealing
discrepancies between the true values and the median estimates (details
about the self-test results are provided in the following section).

\subsection{\texorpdfstring{Robustness of the model to observation error
\label{sec:self-test}}{Robustness of the model to observation error }}\label{robustness-of-the-model-to-observation-error-1}

\begin{figure}

{\centering \includegraphics[width=0.95\linewidth]{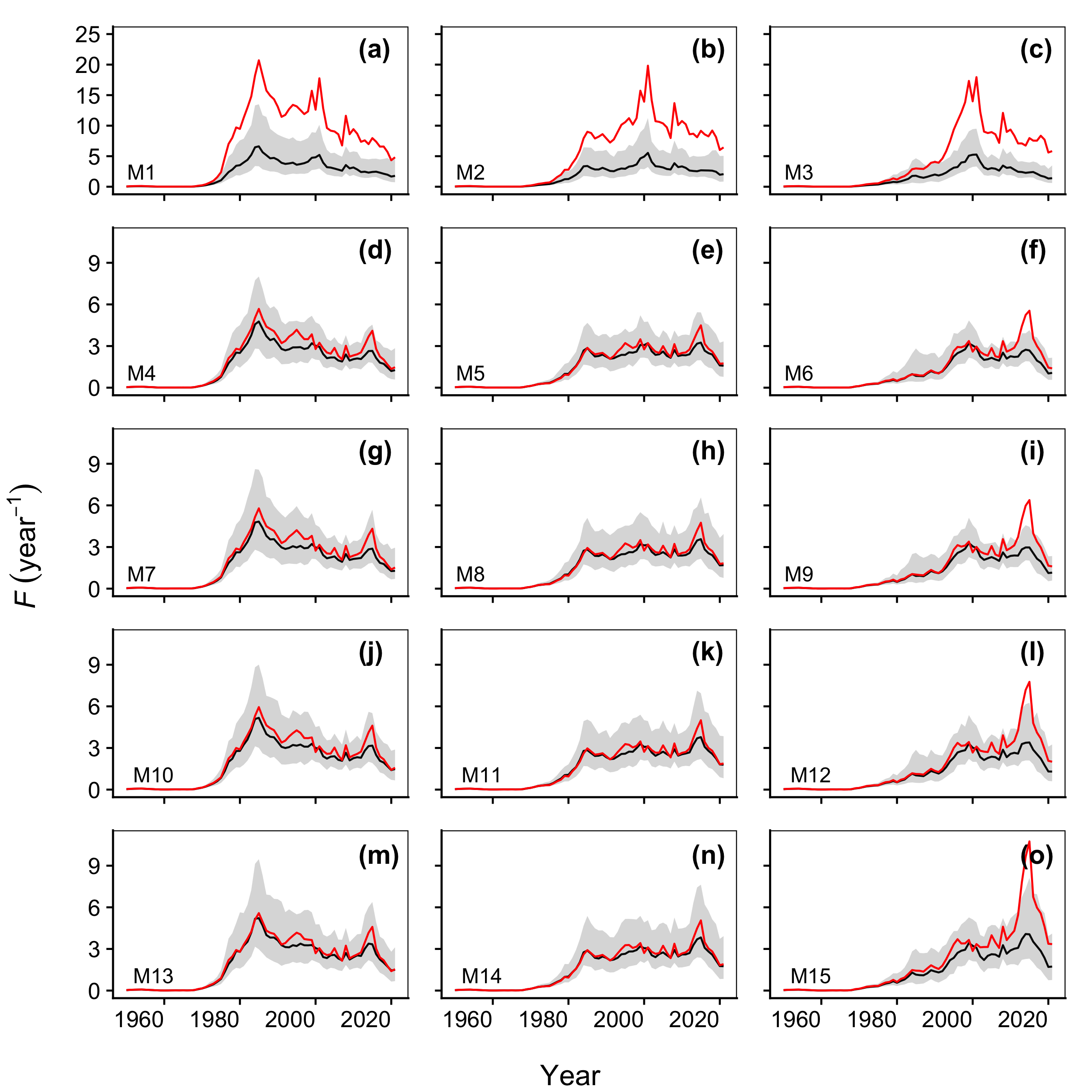} 

}

\caption{Self-test estimates of the fishing mortality rate ($F$) and corresponding true values for all alternative models (M1-15), with the name of the corresponding model denoted at the bottom left corner of each panel. The black lines indicate the median values of the self-test estimates, and the gray areas indicate the 95\% intervals. The red lines indicate the true values. \label{MackSelfF}}\label{fig:unnamed-chunk-14}
\end{figure}

\begin{figure}

{\centering \includegraphics[width=0.95\linewidth]{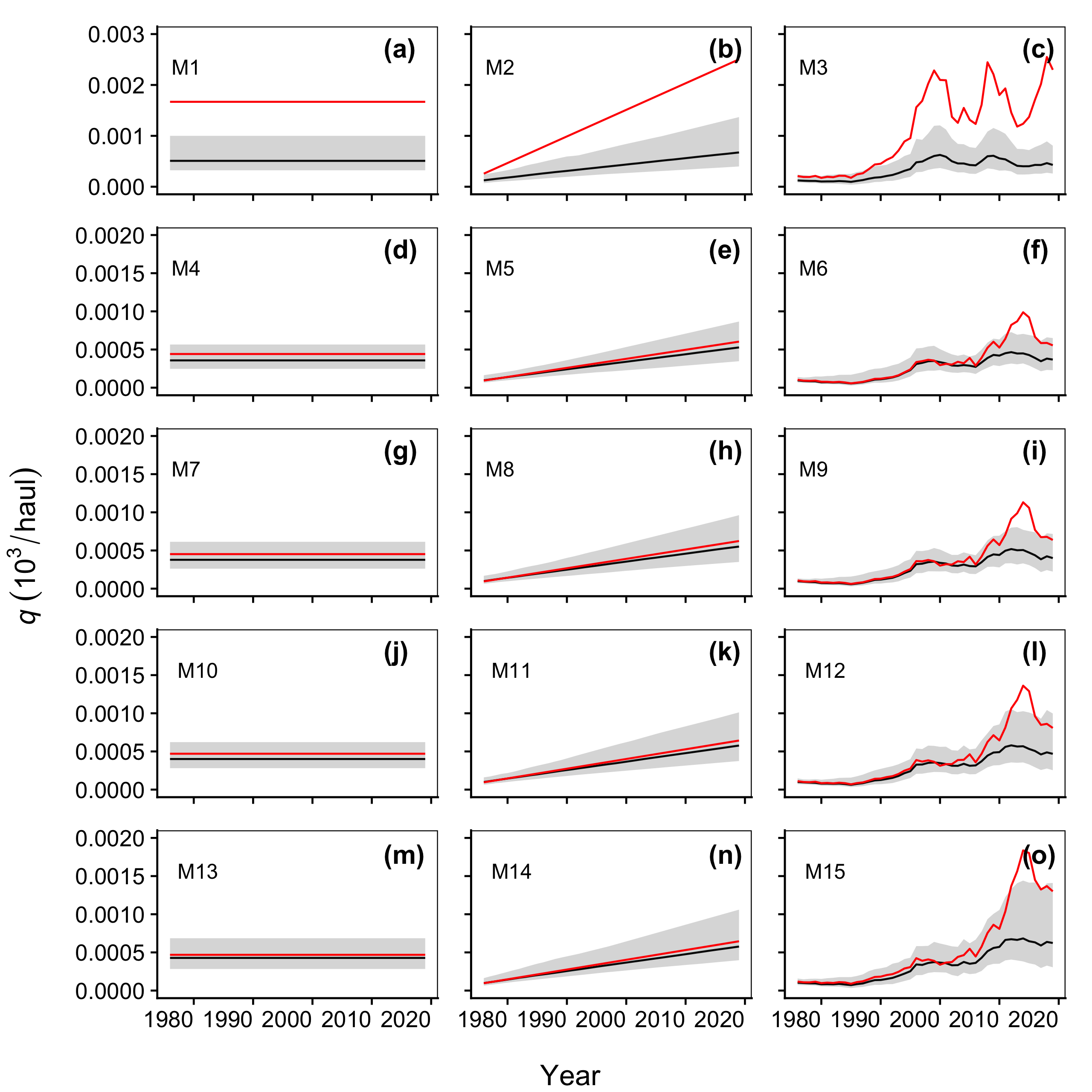} 

}

\caption{Self-test estimates of the catchability ($q$) and corresponding true values for all alternative models (M1-15), with the name of the corresponding model denoted at the top left corner of each panel. The black lines indicate the median values of the self-test estimates, and the gray areas indicate the 95\% intervals. The red lines indicate the true values. \label{MackSelfq}}\label{fig:unnamed-chunk-15}
\end{figure}

The self-test results indicated that the models with time-invariant
selectivity (M1-3) were less robust to observation errors compared to
the models with time-varying selectivity (M4-15) (Figs. \ref{MackSelfF}
and \ref{MackSelfq}). Specifically, the true values consistently fell
outside the 95\% interval of the estimated values for fishing mortality
and catchability time-series in most years when time-invariant
selectivity was assumed (M1-3; see the panels in the first row in Figs.
\ref{MackSelfF} and \ref{MackSelfq}), irrespective of the catchability
assumption. However, all the models, with the exception of M3,
demonstrated a high level of robustness in the SSB time-series (Fig.
\ref{MackSelfSSB}; also in the total biomass time-series; see
Supplementary Fig. \ref{MackSelfB}). Only M3 displayed some deviation
between the true values and the median estimates in the SSB time-series
in the last five years (Fig. \ref{MackSelfSSB}c).

On the contrary, the models with time-varying selectivity (M4-15)
displayed a higher level of robustness against observation errors when
compared to the models with time-invariant selectivity (M1-3) (Figs.
\ref{MackSelfF} and \ref{MackSelfq}). This improved robustness was
evident through the relatively smaller discrepancies between the true
values and the median estimates in the fishing mortality and
catchability time-series. It is worth noting, however, that models with
random walk catchability (M6, M9, M12, M15) exhibited divergent patterns
in the fishing mortality and catchability time-series, especially over
the last decade (see the subfigures f,i,l and o in Figs. \ref{MackSelfF}
and \ref{MackSelfq}). In contrast, models featuring constant
catchability and linearly increasing catchability (M4, M5, M7, M8, M10,
M11, M13, and M14) demonstrated greater robustness across all the
considered time-series quantities (see the subfigures d, e, g, h, j, k,
m, and n in Figs. \ref{MackSelfF}-\ref{MackSelfSSB}).

The convergence rates of the self-test results are summarized in Table
\ref{table:models_config}. In general, the convergence rates were good,
with eight models showing over 90\% convergence out of 500
simulation-estimation runs. However, the models with the highest
auto-correlation (\(\varrho=0.9\)) for selectivity deviations displayed
lower convergence rates, ranging from 51\% to 84\%.

\subsection{Bias and coverage probability of the parameter
estimates}\label{bias-and-coverage-probability-of-the-parameter-estimates}

The RD measures obtained from the parametric bootstrap analysis revealed
that the medians of RDs (i.e., \(\text{median[RD}(\Theta)]\)) were
closely centered around zero, with most of them exhibiting biases within
5\%, except for \(\sigma_{S}\) which showed the largest bias in all 15
models (ranging from \(-9.9\)\% in M5 to \(-21.5\)\% in M15; see
Supplementary Fig. \ref{MackBoot}a-c). Such negative biases in process
error variance parameters were commonly observed in various state-space
stock assessment models
\citep{cadigan2015state, miller2016state, miller2018evaluating, trijoulet2020performance, kim2022state}.
This phenomenon was attributed to the downward bias associated with
maximum likelihood estimation \citep{pawitan2001all}.

Conversely, the SD parameters for the observation error (i.e.,
\(\sigma_{Y}\) and \(\sigma_{I}\)) exhibited slightly positive biases in
the medians of the RDs, which was likely due to the gamma penalties
(compare the medians of the RDs in Supplementary Fig. \ref{MackBoot}).
These positive biases occurred because the estimator shifted the
estimates that were close to zero to slightly larger values, thereby
enhancing parameter identifiability \citep{chung2013nondegenerate}
(Figs. \ref{MackSigY} and \ref{MackSigI}). These biases ranged from
5.3\% to 8.2\% for models with constant catchability (M4-8) and linearly
increasing catchability (M10-14), but were slightly larger for models
with random walk catchability (M9, M12, M15), ranging from 10.1\% to
14.3\%. The biases for \(\sigma_{I}\) were small in all models, ranging
from 0.1\% to 6.0\%. The largest bias for \(\sigma_{I}\) was observed in
M6 (random walk catchability).

The convergence rates of the parametric bootstrap analysis are
summarized in Table \ref{table:models_config}. Overall, the convergence
rates were good, with 10 models showing over 95\% convergence rates out
of 1000 simulation-estimation runs. However, the models with the highest
auto-correlation (\(\varrho=0.9\)) for selectivity deviations displayed
lower convergence rates, ranging from 80\% to 86\%, which aligns with
the tendencies observed in the convergence rates obtained from the
self-test analysis.

The 95\% coverage probabilities of parameter estimates are summarized in
Supplementary Table \ref{table:coverage}, where most coverage
probabilities of the parameters closely aligned with the nominal level
of 95\%. Even those with relatively poor coverage probabilities mostly
fell within the range of 90\% to 98\%. Notably, \(a_{05}\) exhibited
poor coverage probabilities of 82\%, 80\%, and 81\% in M13, M14, and
M15, respectively, possibly attributed to the negative biases of the
point estimates (Supplementary Fig. \ref{MackBoot}a-c). Similarly, the
coverage probabilities for \(\sigma_{Y}\) were also suboptimal in some
models (M6, M9, M12, M13, M15), showing slightly less than 90\% coverage
probability, possibly due to the positive biases of the point estimates.
Models with constant catchability and linearly increasing catchability
demonstrated consistently good coverage probabilities for all
parameters. Small deviations from the nominal level may be attributed to
the use of Wald intervals (i.e., \(\pm 1.96 \cdot \text{SD}\) for 95\%
confidence intervals) for computational efficiency.

\subsection{Estimability of the standard deviation
parameters}\label{estimability-of-the-standard-deviation-parameters}

\begin{figure}

{\centering \includegraphics[width=0.95\linewidth]{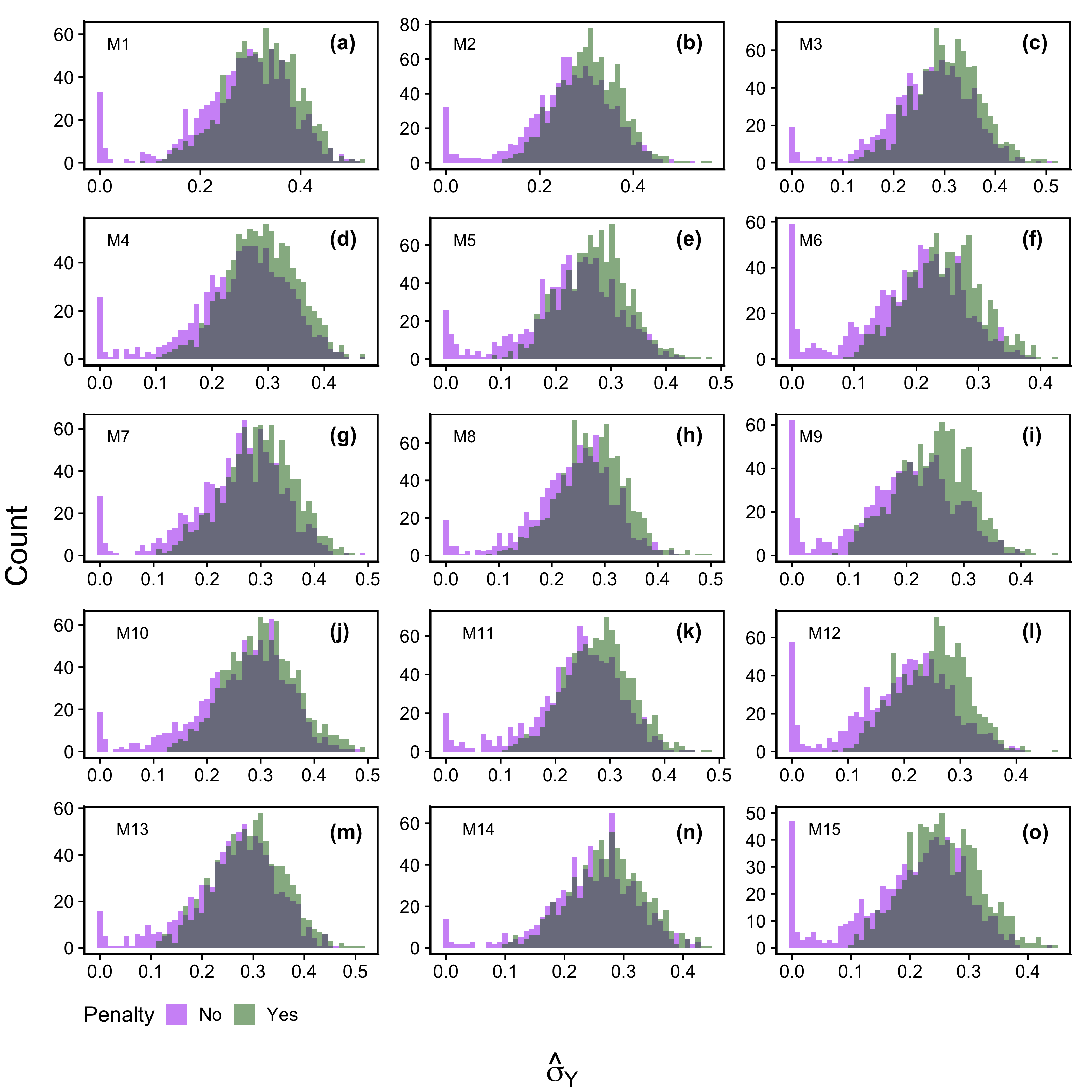} 

}

\caption{Histograms of the estimates of the standard deviation parameter for the observation error of the annual catch ($\widehat{\sigma}_{Y}$), with (dark green) and without (purple) the gamma penalty on the parameter. Each panel displays histograms of the parameter estimated from a different model, with the model name indicated in the top-left corner. The estimates were obtained from the parametric bootstrap analysis. \label{MackSigY}}\label{fig:unnamed-chunk-19}
\end{figure}

\begin{figure}

{\centering \includegraphics[width=0.95\linewidth]{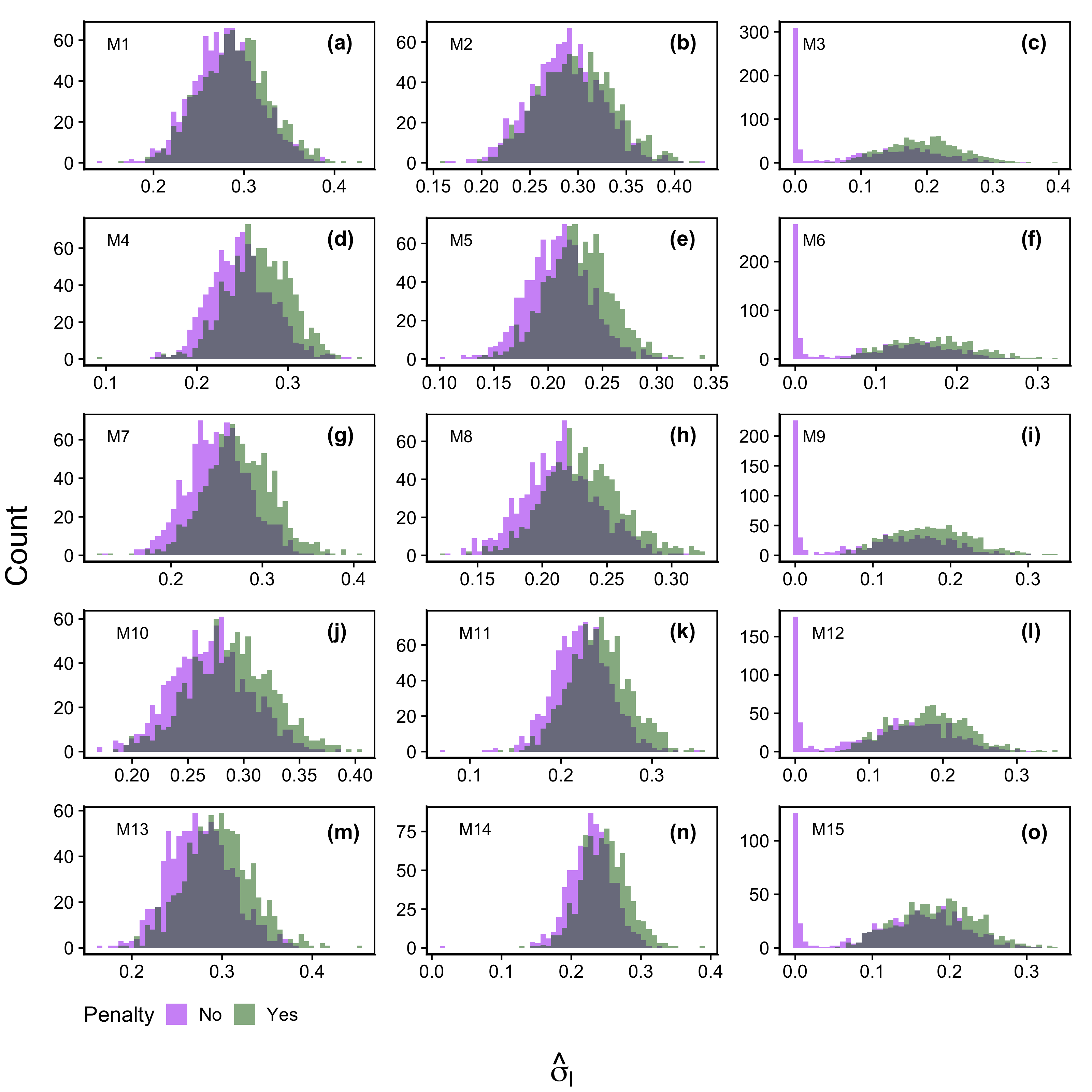} 

}

\caption{Histograms of the estimated values of the standard deviation parameter for the observation error of the CPUE ($\hat{\sigma}_{I}$), with (dark green) and without (purple) the gamma penalty on the parameter. Each panel displays histograms of the parameter estimates from a different model, with the model name indicated in the top-left corner. The estimates were from the parametric bootstrap analysis. \label{MackSigI}}\label{fig:unnamed-chunk-20}
\end{figure}

In cases where gamma penalties were not applied, the distributions of
\(\widehat{\sigma}_{Y}\) in all models, obtained from the parametric
bootstrap analysis, exhibited bimodal shapes, with one mode near zero.
This bimodality suggested that the parameter was not practically
identifiable (Fig. \ref{MackSigY}). Bimodal distributions were also
observed in \(\widehat{\sigma}_{I}\) in models with time-varying
selectivity and random walk catchability (M6, M9, M12, and M15; see Fig.
\ref{MackSigI}). When gamma penalties were applied, both
\(\widehat{\sigma}_{Y}\) and \(\widehat{\sigma}_{I}\) showed unimodal
distributions across all models, indicating that the parameters became
identifiable regardless of the model configurations. However, the
application of gamma penalties slightly shifted the distributions of
\(\widehat{\sigma}_{Y}\) and \(\widehat{\sigma}_{I}\) towards the right,
pushing those estimates that were close to zero towards slightly larger
values (Figs. \ref{MackSigY} and \ref{MackSigI}).

\begin{figure}

{\centering \includegraphics[width=0.95\linewidth]{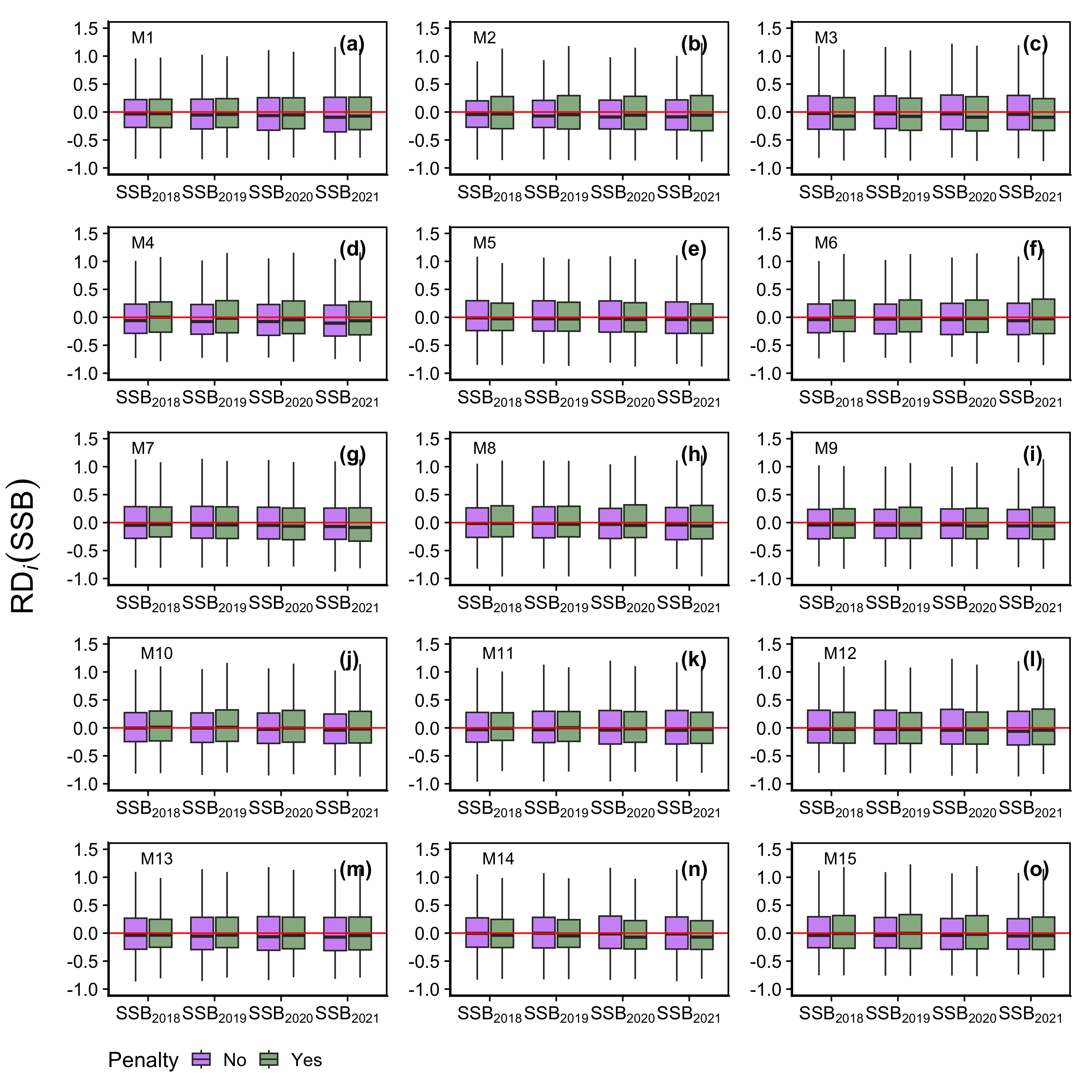} 

}

\caption{Boxplots of the relative differences (RD) of the spawning stock biomass ($SSB$) for the last four years of the time series from all 15 alternative models (M1-15) compared to their corresponding true values. The horizontal red line indicates unbiasedness (i.e., RD = 0). Each panel corresponds to a different model, with the model name indicated in the top-left corner. Each colored boxplot represents the RDs from models with gamma penalties (dark green) and without gamma penalties (purple).  \label{RDSSB}}\label{fig:unnamed-chunk-42}
\end{figure}

\begin{figure}

{\centering \includegraphics[width=0.95\linewidth]{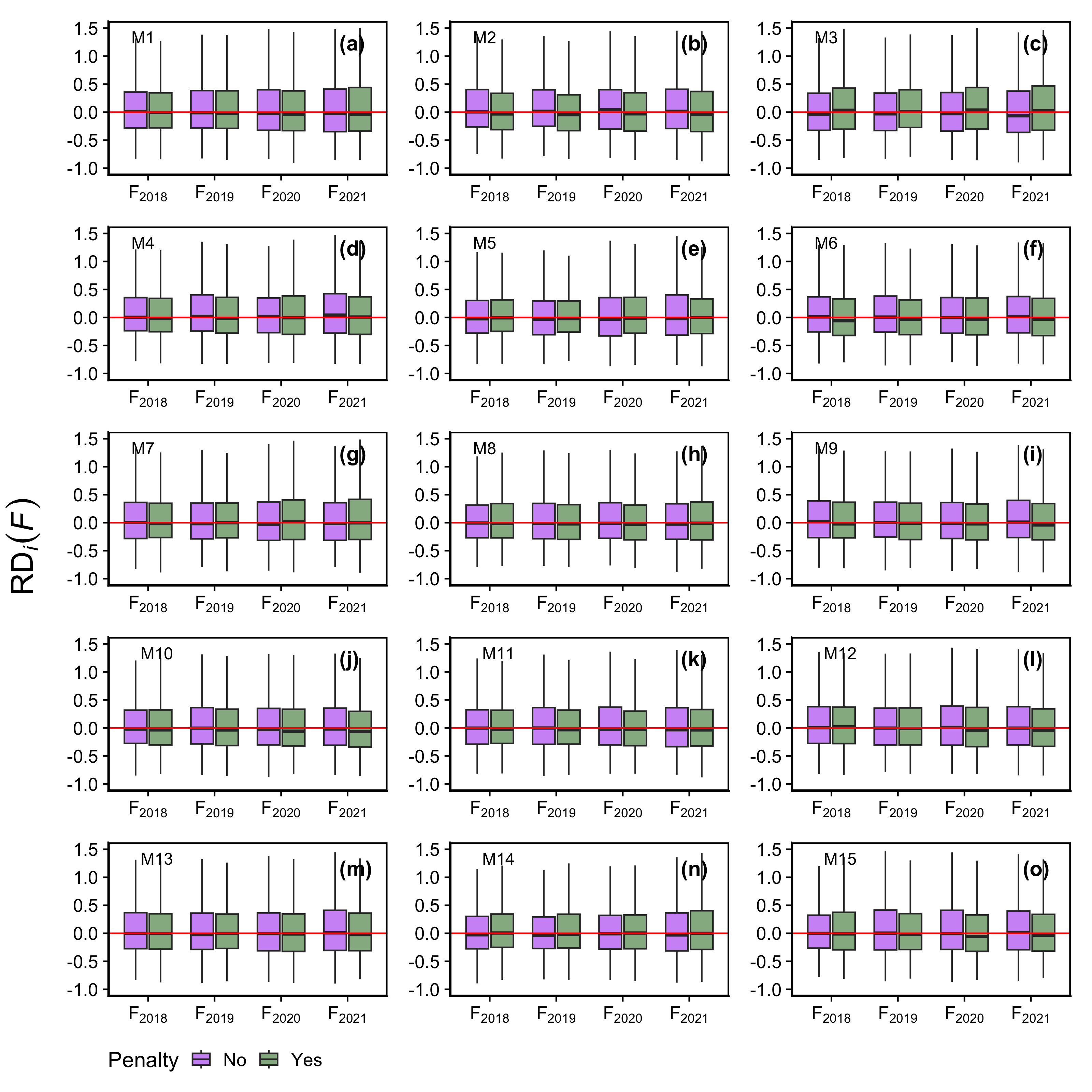} 

}

\caption{Boxplots of the relative differences (RD) of the fishing mortality rate ($F$) for the last four years of the time series from all 15 alternative models (M1-15) compared to their corresponding true values. The horizontal red line indicates unbiasedness (i.e., RD = 0). Each panel corresponds to a different model, with the model name indicated in the top-left corner. Each colored boxplot represents the RDs from models with gamma penalties (dark green) and without gamma penalties (purple). \label{RDF}}\label{fig:unnamed-chunk-43}
\end{figure}

While the penalties significantly improved the estimability of
\(\sigma_{Y}\) and \(\sigma_{I}\), these enhancements did not have a
substantial impact on the estimates of derived time-series quantities,
such as SSB and fishing mortality (Supplementary Figs. \ref{RDSSB} and
\ref{RDF}). This is likely because the other random effects in the model
(e.g., modeled random deviations in selectivity, recruitment, fishing
mortality) had much larger impacts on the derived time-series quantities
than those associated with the observation SD parameters. This
phenomenon was observed in previous studies, where observation error
variance parameters were not practically identifiable, but state
variables and derived time-series quantities were still estimated with
minimal bias \citep{auger2016state, kim2022state}, especially when
observation error variance parameters were smaller than process error
variance parameters \citep{auger2016state, kim2022state}.

\subsection{Validation of the random effects
assumptions}\label{validation-of-the-random-effects-assumptions}

\begin{figure}

{\centering \includegraphics[width=0.95\linewidth]{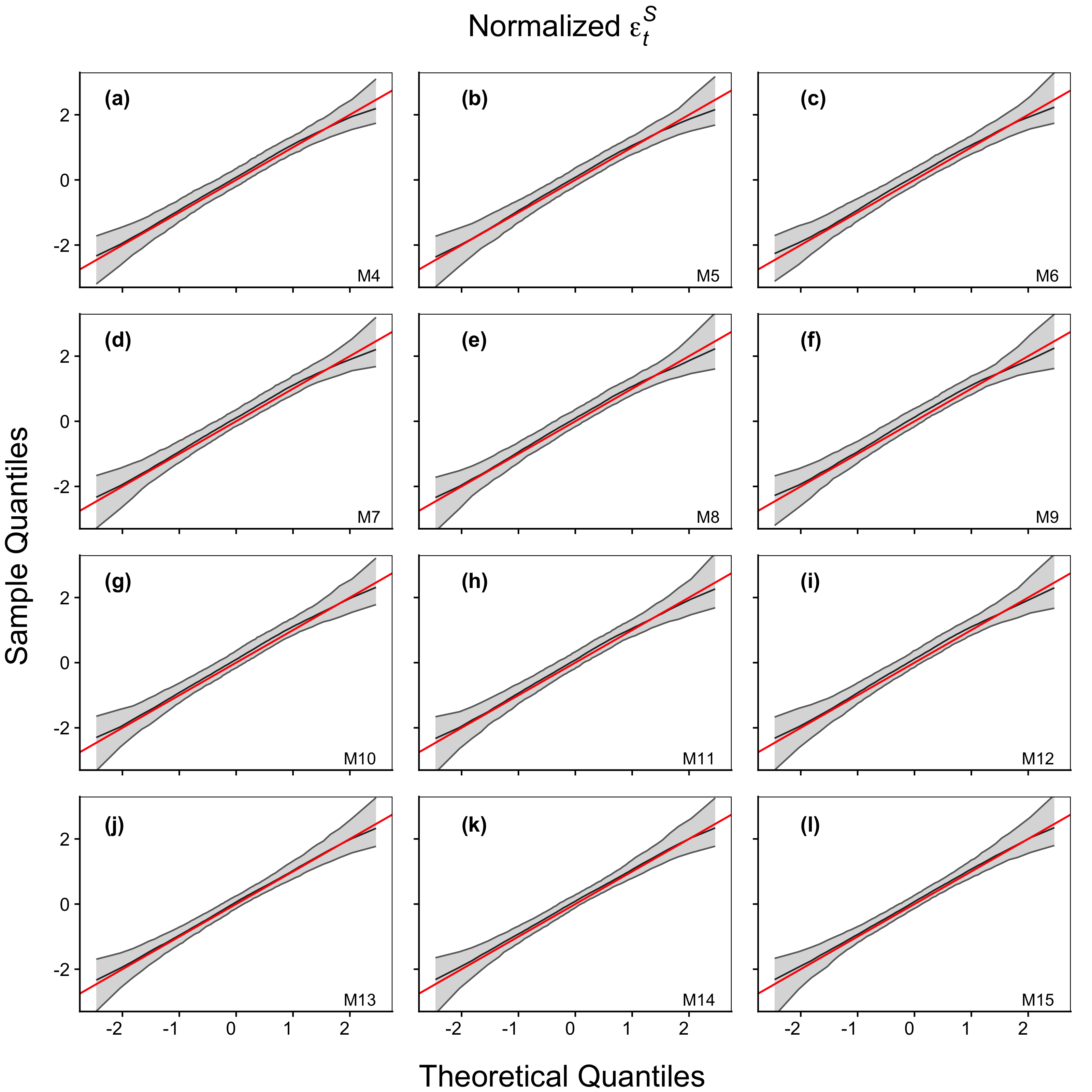} 

}

\caption{Quantile-quantile (Q-Q) plots for the standardized samples of selectivity deviations for the time-varying age at 5\% selectivity ($a_{05, t}$) in models assuming time-varying selectivity (M4-15). Each panel corresponds to a different model, with the model name indicated in the bottom-right corner. The diagonal red line serves as the reference for normality check. The 95\% envelope of the Q-Q plot is shaded in gray, and the black line represents the median of the envelope. \label{SamplesSel}}\label{fig:unnamed-chunk-44}
\end{figure}

\begin{figure}

{\centering \includegraphics[width=0.95\linewidth]{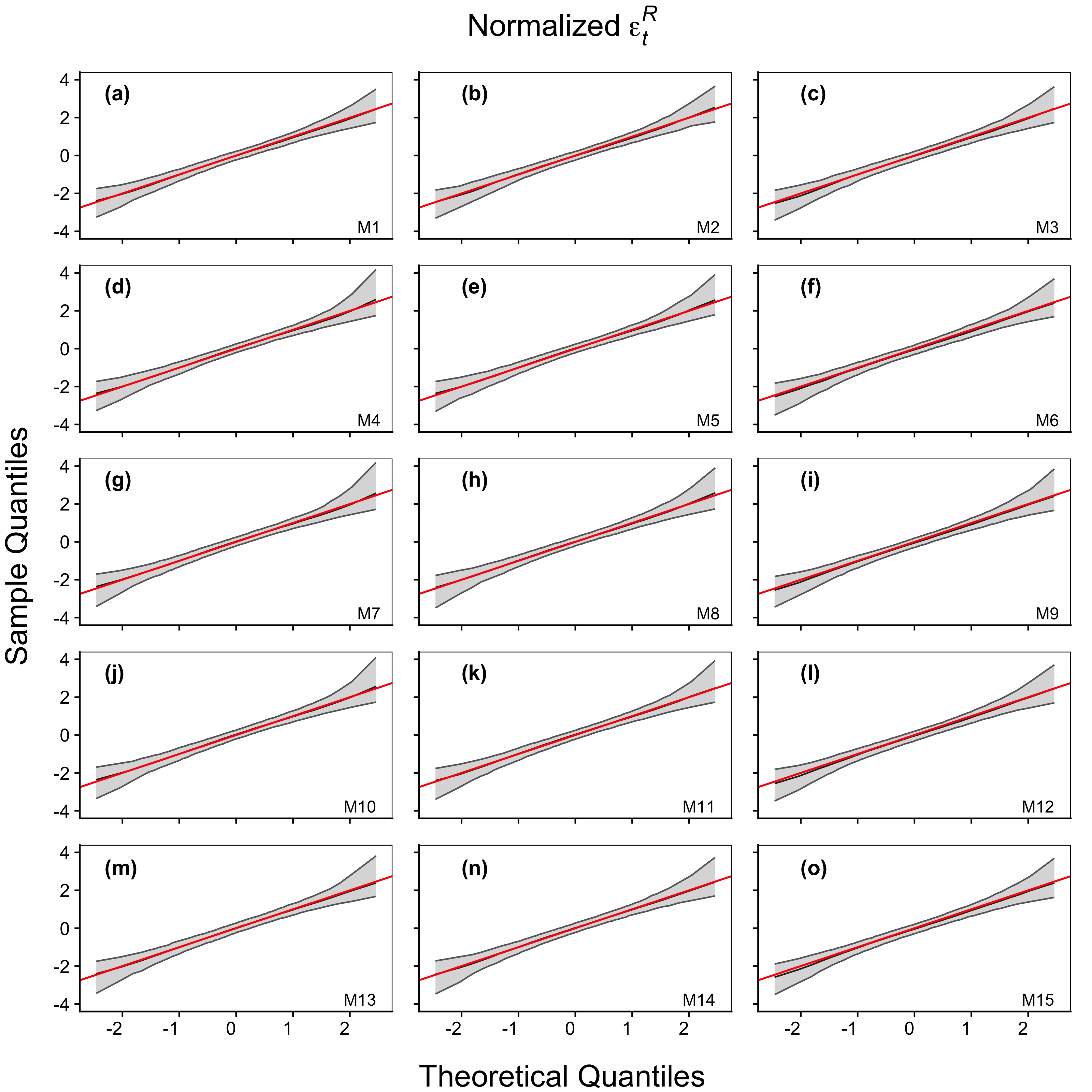} 

}

\caption{Quantile-quantile (Q-Q) plots for the standardized samples of inter-annual deviations of the recruitment deviations from all alternative models (M1-15). Each panel corresponds to a different model, with the model name indicated in the bottom-right corner. The diagonal red line serves as the reference for normality check. The 95\% envelope of the Q-Q plot is shaded in gray, and the black line represents the median of the envelope. \label{SamplesRec}}\label{fig:unnamed-chunk-45}
\end{figure}

\begin{figure}

{\centering \includegraphics[width=0.95\linewidth]{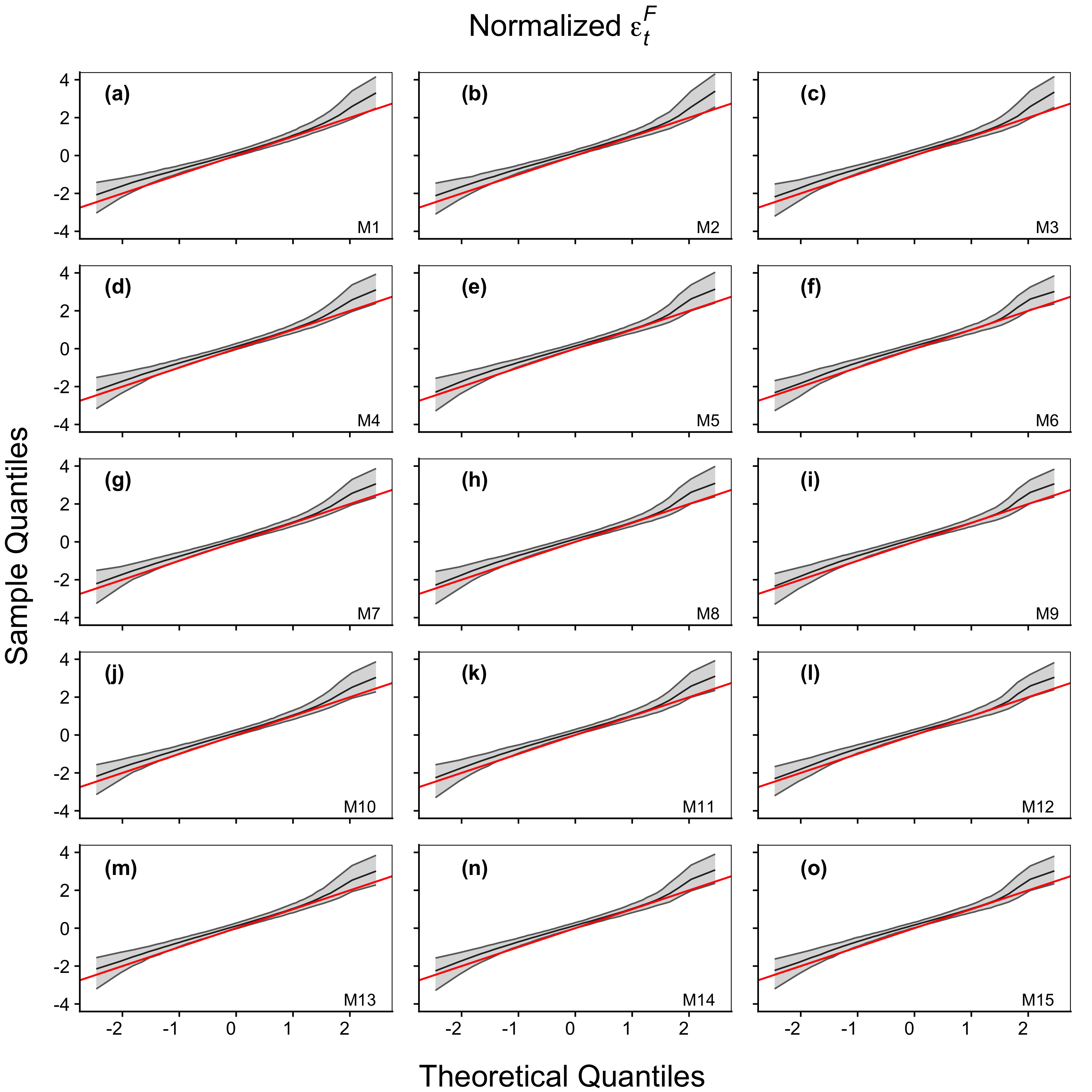} 

}

\caption{Quantile-quantile (Q-Q) plots for the standardized samples of inter-annual deviations of the fishing mortality rates in log-scale from all alternative models (M1-15). Each panel corresponds to a different model, with the model name indicated in the bottom-right corner. The diagonal red line serves as the reference for normality check. The 95\% envelope of the Q-Q plot is shaded in gray, and the black line represents the median of the envelope. \label{SamplesFishing}}\label{fig:unnamed-chunk-46}
\end{figure}

\begin{figure}

{\centering \includegraphics[width=0.95\linewidth]{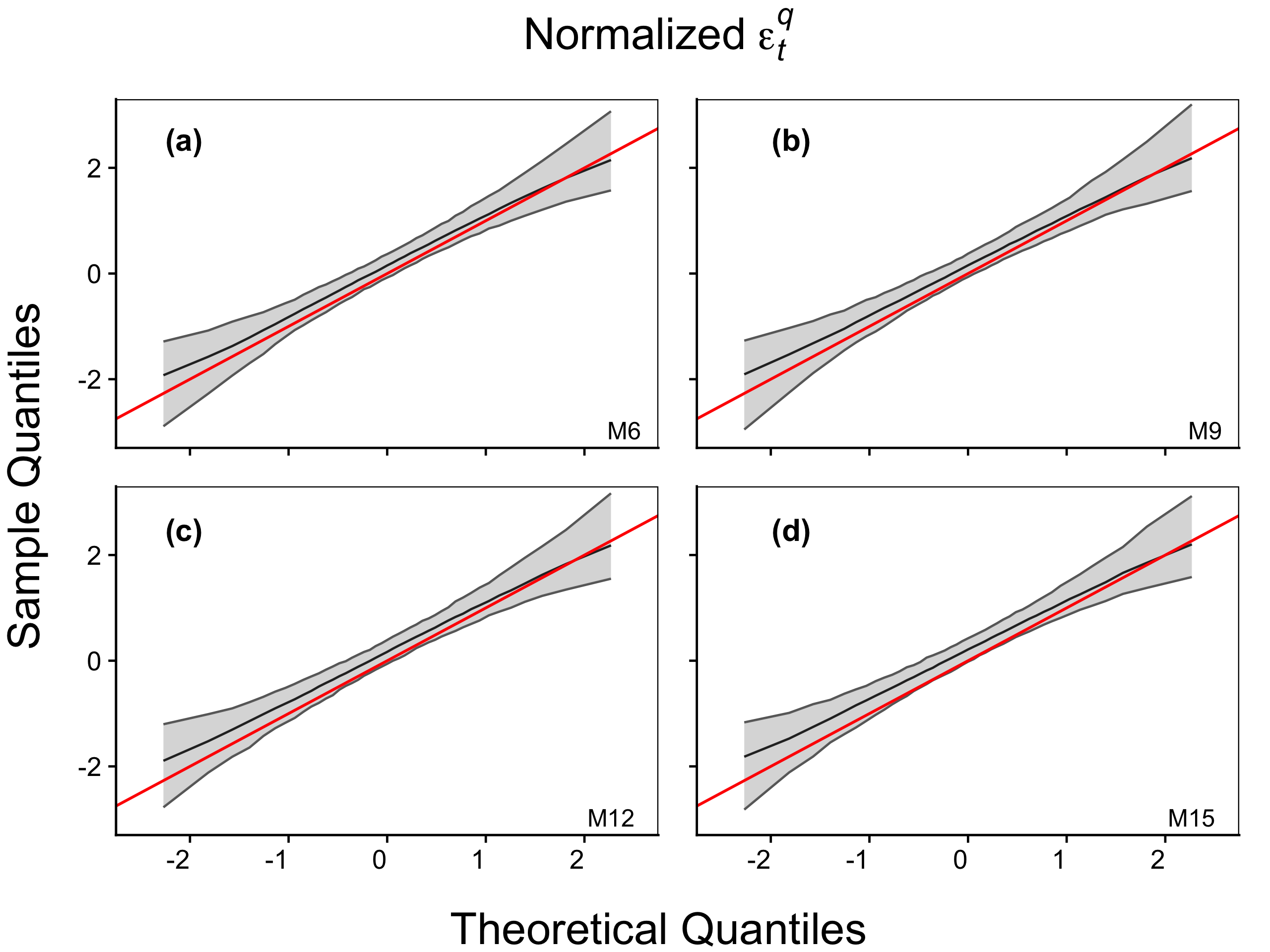} 

}

\caption{Quantile-quantile (Q-Q) plots for the standardized samples of inter-annual deviations of the catchability in log-scale from models assuming random walk catchability (M6, M9, M12, M15). Each panel corresponds to a different model, with the model name indicated in the bottom-right corner. The diagonal red line serves as the reference for normality check. The 95\% envelope of the Q-Q plot is shaded in gray, and the black line represents the median of the envelope. \label{SamplesCatchability}}\label{fig:unnamed-chunk-47}
\end{figure}

The standardized samples of random effects conformed well to the
standard normal distribution across all models. This was evident from
the respective Q-Q plots, where the 95\% envelopes mostly covered the
diagonal lines (Supplementary Figs.
\ref{SamplesSel}-\ref{SamplesCatchability}). However, minor systematic
deviations from the diagonal lines were noted in the right tail of the
median Q-Q lines for fishing mortality deviations (Supplementary Fig.
\ref{SamplesFishing}), and in the left tail for catchability deviations
(Supplementary Fig. \ref{SamplesCatchability}). Despite these
deviations, the median Q-Q lines were generally close to the diagonal
lines, with the 95\% envelopes largely encompassing these diagonals.

\section{Discussion}\label{discussion}

This study demonstrated the efficacy of our proposed model in addressing
unexplained temporal variability in CPUE data, without requiring
additional information. Furthermore, we highlighted the capabilities of
the non-degenerate estimator in handling non-identifiability issues
related to the SD parameters, thus facilitating the simultaneous
estimation of all SD parameters within the model. We recommend adopting
this approach in future stock assessments employing a state-space
framework, especially when it is crucial to estimate both observation
and process error SD parameters. However, we also acknowledge that the
non-degenerate estimator introduces small bias into the estimates of SD
parameters. This bias is likely to be negligible if the true SD is not
too close to zero as demonstrated by \citet{chung2013nondegenerate}, but
it is important to be aware of this bias and evaluate its potential
impacts on stock assessments through simulation studies, such as the
parametric bootstrap test that we conducted in this study. Furthermore,
the inclusion of additional penalty terms for SD parameters in more
complex models rendered AIC less effective for model selection, as
demonstrated in our study. This indicates that the AIC may not be
ideally suited for comparing models that incorporate differing numbers
of penalty terms for SD parameters. Treating the penalty terms for SD
parameters as priors can be an alternative approach to address this
issue while treating SD parameters as random effects, as incorporated in
WHAM \citep{stock2021woods}, but our initial attempts to implement this
approach were not successful due to convergence issues. Nonetheless,
within the scope of our research, the AIC proved valuable in identifying
the most effective model between those with constant and linearly
increasing catchability with time-varying selectivity, since the penalty
terms for SD parameters were consistent across these models.
Importantly, these were the only models that met all criteria in our
model checking tests.

Additionally, our findings highlighted significant enhancements achieved
by incorporating time-varying selectivity and catchability into the
assessment model. This integration has led to a better performing model,
characterized by reduced retrospective patterns and improved
goodness-of-fit. In our study, time-varying selectivity was more
influential in addressing the retrospective patterns than time-varying
catchability, but both components were essential in improving the model
performance in terms of goodness-of-fit. The best-performing model (M11)
indicated the presence of auto-correlation in the age at 5\%
selectivity, suggesting that the temporal changes in the left limb of
the length composition data (temporal variations in the proportion of
small individuals) were partially explained by changes in selectivity,
not solely by recruitment signals. However, care should be taken when
interpreting this result, as we only investigated the relationship
between the auto-correlation in selectivity deviations and the temporal
changes in the length composition data. This relationship may be
influenced by the auto-correlation in recruitment deviations, which we
did not investigate in this study. We recommend that future studies
explore these potential correlations to more thoroughly examine the
underlying mechanisms of the temporal changes in the distribution of the
length composition data. For example, a two-dimensional AR(1) process,
applied in previous studies
\citep{cadigan2015state, xu2019new, stock2021woods}, can be used to
investigate the correlation between recruitment and selectivity
deviations and their auto-correlations over time.

We demonstrated that in cases where CPUE standardization is not
feasible, our model-based approach provides a valuable alternative. This
approach implicitly accounts for variations in CPUE without the need for
supplementary data typically required by explicit CPUE standardization
methods. However, it is crucial to exercise caution when implementing
this approach using a random walk, as the flexibility of the random walk
process may lead to overfitting and robustness issues. In our study, the
random walk process for catchability was not effective in addressing the
retrospective patterns, and it showed divergent patterns in the
self-test (note that we refer to the self-test as the test where the
observation errors were randomly generated while fixing the process
errors, whereas the parametric bootstrap test refers to the test where
both the observation and process errors were randomly generated). This
is likely due to the fact that the random walk process was overly
flexible and sensitive to the few data points influencing the overall
trend. To mitigate these challenges and enhance model performance, we
recommend exploring alternative formulations for the time-varying
components, such as the simple linear trend that we used for
time-varying catchability.

Furthermore, our self-test and parametric bootstrap results unveiled
numerous potential issues with the model, encompassing biased estimates,
poor convergence, and non-identifiability challenges. These
simulation-based diagnostics have effectively identified misspecified
models. This highlighted the importance of conducting simulation studies
to thoroughly assess and ensure model performance and robustness,
aligning with recommendations from previous studies
\citep{deroba2015simulation, auger2016state, auger2021guide, kim2022state}.
The full randomization of both the observation and process errors in the
parametric bootstrap test was particularly useful in investigating
non-identifiable parameters as well as checking uncertainty estimates
through the coverage test, but this full simulation-estimation test did
not indicate misspecification of the model to the observed data, as the
data were simulated based on the true realizations of the process errors
in each simulation run. In contrast, the self-test, where the process
errors were fixed at the fitted values as the true values, and only the
observation errors were randomized, was effective in identifying
misspecified models, but it did not indicate issues with
non-identifiability because the test was not capable of investigating
the separability of the observation and process errors. Therefore, we
recommend conducting both self-test and parametric bootstrap to broadly
assess model performance and robustness, especially when the model is
complex and the data are sparse.

In the context of stock assessment results for the mackerel stock, our
study highlighted the potential for significant improvements in future
assessments, particularly through the incorporation of models that
consider fishery-dependent processes and the accurate utilization of the
sparse age-length information. Notably, previous assessments of this
stock overlooked the impacts of fishery-dependent processes on the CPUE
data
\citep{jung2019bayesian, gim2020management, jung2021bayesian, hong2022stock, gim2022application, kim2022state, gim2023assumptions}
and misapplied the age-length key when using age-structured models for
the length frequencies \citep{gim2020management, gim2023assumptions}.
These simplifications and misapplications could introduce bias into
assessments of stock status and fishing mortality rates, potentially
leading to poor management decisions. In fact, our assessments differed
from previous ones, showing more pessimistic results. This difference
occurred as some positive signals from the CPUE and length composition
data were partially offset by the inclusion of time-varying catchability
and selectivity. This underscores the importance of conducting thorough
investigations into the potential impacts of fishery-dependent processes
on key management quantities. Such examination is essential to ensure
that assessment results are not unduly influenced by these processes.
For example, our best model (M11), where the linear increase in
catchability was included, supported the presence of vessel power creep
of the LPS fishery, which was reported in the previous study
\citep{seo2017change}. If future assessments are to incorporate this
factor, the TAC advice may be more conservative than that from previous
assessments, which did not account for the vessel power creep. Other
catchability trends, such as asymptotic or exponential increases, may
also be considered in future assessments, as these trends may better
capture the underlying mechanisms of the catchability changes. Moreover,
investigating the impact of including these time-varying components on
projections may be required for future assessments and management
advice, which we recommend as an area for future study.

To the best of our knowledge, this study is the first that incorporated
fishery-dependent processes through time-varying catchability and
selectivity into the assessment of the Korean mackerel stock, as well as
conducted thorough model checking and simulation-based diagnostics to
assess model performance and robustness of the results. We believe that
our findings will provide valuable insights into future assessments of
this stock, but we also acknowledge that our results may not represent
the current status of the stock. This is because our assessment model
was not developed with the most up-to-date data due to data availability
and required simplifying assumptions on biological parameters, which are
often necessary for data-limited assessments. For example, estimating
length-at-age distributions in catch using the samples collected from
the single year (i.e., year 2015) overlooks potential changes in growth
of the species over time. If more data are available, a segmented
approach, estimating the length-at-age distributions in distinct time
blocks, can be considered to account for potential changes in growth
over time. Furthermore, the selection of the DM distribution for models
with time-varying selectivity warrants additional scrutiny. Although
endorsed by \cite{xu2020comparing}, a study by \cite{fisch2021assessing}
indicated that a logistic normal distribution outperformed the DM
distribution for length composition data with large sample sizes.
Despite these limitations, regarding model development and diagnostics,
we believe that our study laid a solid foundation for a better
understanding of the Korean mackerel stock and the development of future
assessment models.

\section*{Acknowledgments}\label{acknowledgments}
\addcontentsline{toc}{section}{Acknowledgments}

We express our gratitude to the marine science team at Dragonfly Data
Science for their valuable discussions and manuscript review. Special
thanks to Dr.~Philipp Neubauer for his insightful comments and project
management support. We also extend our appreciation to the associate
editor and two anonymous reviewers for their constructive feedback and
suggestions.

\section*{Author contribution
statement}\label{author-contribution-statement}
\addcontentsline{toc}{section}{Author contribution statement}

\textbf{Kyuhan Kim:} Conceptualization, Data curation, Formal analysis,
Investigation, Methodology, Software, Validation, Visualization, Writing
-- original draft, Writing -- review \& editing.

\textbf{Nokuthaba Sibanda:} Conceptualization, Supervision, Writing -
review \& editing.

\textbf{Richard Arnold:} Conceptualization, Supervision, Writing -
review \& editing.

\textbf{Teresa A'mar:} Conceptualization, Supervision, Writing - review
\& editing.

\section*{Funding statement}\label{funding-statement}
\addcontentsline{toc}{section}{Funding statement}

Funding: Kyuhan Kim and Teresa A'mar were supported by the Dragonfly
internal marine R\&D project.

\section*{Data availability
statement}\label{data-availability-statement}
\addcontentsline{toc}{section}{Data availability statement}

Data and code to reproduce the results in this paper are available in
the Zenodo repository (10.5281/zenodo.11177291,
\url{https://doi.org/10.5281/zenodo.11177291}).

\section*{Competing interests}\label{competing-interests}
\addcontentsline{toc}{section}{Competing interests}

The authors declare there are no competing interests.

\bibliography{references}

\pagebreak

\renewcommand{\arraystretch}{0.75}
\begin{table}[]
\centering
\caption{Definition of terms used in this study} \label{table:terms}
\begin{tabular}{ l l }
\hline
 Notation & Description \\ 
 \hline
 $t$, $a$, $j$ & Indices for year, age, and length bin, respectively \\
 $R$ & Average recruitment \\
 $\varepsilon^{R}_{t}$ & Inter-annual deviations in natural logarithm of average recruitment \\ 
 $A$ & Terminal age class \\
 $M$ & Instantaneous rate of natural mortality \\
 $N_{a,t}$ & Abundance by age and year \\
 $\widehat{C}_{a,t}$ & Predicted catch by age and year \\
 $\widehat{C}_{j,t}$ & Predicted catch by length group and year \\
 $S_{a,t}$ & Fishery selectivity by age and year \\
 $Z_{a,t}$ & Instantaneous rate of total mortality by age and year \\
 $F_{a,t}$ & Instantaneous rate of fishing mortality by age and year  \\
 $F_{0}$ & Average instantaneous rate of fully selected fishing mortality for a pre-data period \\
 $F_{t}$   & Instantaneous rate of fully selected fishing mortality by year  \\
 $\varepsilon^{F}_{t}$ & Inter-annual deviations in natural logarithm of fishing mortality \\ 
 $\sigma^{2}_{F}$ & Variance of inter-annual deviations in natural logarithm of fishing mortality \\
 $a_{95}$ & Age at 95\% selectivity \\
 $a_{05}$ & Age at 5\% selectivity used as a constant or median value for a time-varying model \\
 $l_{\Theta}, u_{\Theta}$ & Lower and upper bounds for a bounded logit transformation of a parameter $\Theta$ \\
 $a_{05,t}$ & Time-varying age at 5\% selectivity \\
 $\varepsilon^{S}_{t}$ & Inter-annual deviations in logit-transformed age at 5\% selectivity \\
 $\boldsymbol{\varepsilon}^{S}$ & Vector of inter-annual deviations in logit-transformed age at 5\% selectivity \\
 $\sigma^{2}_{S}$ &  Variance of inter-annual deviations in logit-transformed age at 5\% selectivity \\
 $\varrho$ &  Autocorrelation coefficient for $\varepsilon^{S}_{t}$ in a stationary AR(1) process \\
 $q_{0}$ & Catchability coefficient used as a constant or initial value for a time-varying model \\
 $q_{t}$ & Time-varying catchability coefficient \\
 $c$ & Additive constant for the linear increase in catchability \\
 $\varepsilon^{q}_{t}$ & Inter-annual deviations in catchability \\
 $\sigma^{2}_{q}$ & Variance of inter-annual deviations in catchability \\
 $\pi_{j|a}$ &  Probability of fish of age $a$ being in length group $j$ \\
 $\Phi(\cdot)$ & Cumulative distribution function of a skew normal distribution \\
 $\xi_{a}$, $\omega_{a}$, $\alpha_{a}$ &  Location, scale, and shape parameters of a skew normal distribution \\
 $I_{t}$, $\widehat{I}_{t}$ &  Observed and predicted catch-per-unit-effort (CPUE), respectively \\
 $\sigma^{2}_{I}$ &  Variance of observation errors in CPUE \\
 $\text{VB}_{t}$ & Vulnerable biomass by year \\
 $W_{a}$ & Mean weight-at-age \\
 $L_{a}$ & Mean length-at-age \\
 $\text{SSB}_{t}$ & Spawning stock biomass by year \\
 $m_{a}$ & Mean maturity-at-age \\
 $\phi$ & Fraction of a year elapsed when spawning \\
 $\varphi$ & Female proportion in a population \\
 $Y_{t}$, $\widehat{Y}_{t}$ & Observed and predicted catch (in weight) by year, respectively \\
 $\sigma^{2}_{Y}$ &  Variance of observation errors in catch (in weight) \\
 $\boldsymbol{n}_{t}$ &  Vector of observed length frequencies by year \\
 $E_{t}$ & Sample size of observed length frequencies by year \\
 $\delta_{t}$ & Concentration parameters of the Dirichlet-multinomial distribution \\
 $\boldsymbol{\widehat{P}}_{t}$ & Vector of predicted length proportions by year \\ 
 $\widehat{P}_{j|t}$ &  Predicted proportion of fish in length group $j$ by year \\
 $E^{\text{eff}}_{t}$ & Effective sample size of observed length frequencies by year \\
 $\theta$ & Scale parameter in the Dirichlet-multinomial (DM) distribution \\
 $\delta_{j,t}$, $\beta_{t}$ & Concentration and variance-inflation parameters in the DM distribution, respectively \\
 $\boldsymbol{\Theta}$, $\boldsymbol{Q}$, $\boldsymbol{D}$ & Vector of all parameters, random effects, and data, respectively \\
 $\psi$,$\lambda$ & Shape and rate parameters of the gamma distribution, respectively \\
  \hline
\end{tabular}
\end{table}

\clearpage

\renewcommand{\arraystretch}{1.2}
\begin{table}[]
\centering
\caption{Description of quantities and sub-models with corresponding parameter values and references. The minimum, maximum, and mean lengths were utilized in estimating the length-at-age distributions in catch using skew normal distributions. In cases where the column is left blank under the "Reference" heading, it denotes that the value was not obtained from a published source and instead chosen based on the best guess of the authors of this study.}
\label{table:submodels}
\begin{tabular}{ l l l }
\hline
 Description (notation) & Quantity or Equation & Reference \\ 
 \hline
  $\bf{Quantity}$                       &                                      &                      \\
  Minimum length-at-age                 & $19.1, 23.1, 26.1, 30.1, 31.1, 35.1$ (cm) & \citet{jung2021study} \\
  Maximum length-at-age                 & $29.0, 33.0, 36.0, 42.0, 43.0, 44.0$ (cm) & \citet{jung2021study} \\
  Mean length-at-age ($L_{a}$)          & $23.9, 27.8, 30.6, 34.8, 37.2, 40.1$ (cm) & \citet{jung2021study} \\
  Natural mortality ($M$)               & $0.53$ (year$^{-1}$) & \citet{nishijima2021update}  \\
  Year elapsed when spawning ($\phi$)    & $0.5$ (fraction of the year)  & \citet{kim2020maturity}   \\
  Female proportion ($\varphi$)         & $0.5$  &   \\
  Pre-data fishing mortality ($F_{0}$) & $0.01$ (year$^{-1}$)  &  \\ 
  $\bf{Model}$                       &                                      &                      \\
  Mean weight-at-age ($W_{a}$)    & $W_{a}= 0.003 \cdot 10^{-6} \cdot L_{a}^{3.425}$ (MT)   & \citet{gim2019importance} \\
  Mean maturity-at-age ($m_{a}$)      & $m_{a}=1/[1+\exp(20.11- 0.7 \cdot L_{a} )] $ & \citet{kim2020maturity}  \\
 \hline
\end{tabular}
\end{table}

\clearpage

\begin{table}[]
\centering
\caption{ Configuration of all 15 models fitted to the mackerel data, based on various assumptions concerning catchability ($q_{t}$), selectivity ($a_{05,t}$), and autocorrelation ($\varrho$) of the selectivity deviation. $\Delta$AIC represents the difference in AIC compared to the best model (M11). Convergence rates denote the percentage of successful convergence out of 500 for self-tests with randomized observation errors (Self-obs) and 1000 for self-tests with both randomized observation and process errors (Self-both; parametric bootstrap).} \label{table:models_config}
\begin{tabular}{ l l r r r r r }
\hline   
\multirow{2}{*}{Model} & \multicolumn{3}{c}{Assumption} & \multirow{2}{*}{$\Delta$AIC} & \multicolumn{2}{c}{Convergence rate (\%)} \\ \cline{2-4} \cline{6-7} 
     & $q_{t}$ & $a_{05,t}$ & $\varrho$ &  & Self-obs & Self-both  \\
 \hline
 M1    & Constant & Constant        & \textemdash & $112.914$ & 89 & 97  \\
 M2    & Linear increase & Constant  & \textemdash & $101.948$ & 94 & 95                    \\
 M3    & Random walk & Constant     & \textemdash & $197.440$ & 79 & 96     \\
 M4    & Constant & Random & 0.0 & $30.656$ & 96 & 92 \\
 M5    & Linear increase & Random & 0.0 & $6.488$ & 96 & 96 \\
 M6    & Random walk & Random & 0.0  & $99.117$ & 95 & 86 \\
 M7    & Constant & Random & 0.3 & $26.209$ & 90 & 96 \\
 M8    & Linear increase & Random  & 0.3 & $1.920$ & 96 & 96 \\
 M9    & Random walk & Random & 0.3 & $93.658$ & 86 & 95 \\
 M10   & Constant & Random & 0.6 & $24.904$  & 91 & 95 \\
 M11   & Linear increase & Random & 0.6 & $0.000$ & 95 & 96 \\
 M12   & Random walk & Random & 0.6 & $90.512$ & 84 & 95 \\
 M13   & Constant & Random & 0.9 & $25.508$ & 51 & 86 \\
 M14   & Linear increase & Random & 0.9 & $2.683$ & 84 & 80 \\
 M15   & Random walk & Random & 0.9  & $92.585$ & 67 & 80 \\
 \hline
\end{tabular}
\end{table}

\clearpage

\begin{table}[]
\centering
\caption{Equations for the joint likelihood of all alternative models (M1-15) when fitted, along with components of the likelihood for observation error and process error. Gamma distributions were applied as penalty functions for three standard deviation parameters ($\sigma_{I}$, $\sigma_{Y}$, and $\sigma_{q}$). Note that the gamma distribution here is parameterized by the shape and rate (i.e., the reciprocal of the scale parameter) parameters.} \label{table:models_likelihood}
\begin{tabular}{ l l }
\hline
 Description & Equation   \\ 
 \hline
 \bf{Likelihood (process error)} \\
  Recruitment deviation & $\begin{aligned} \mathcal{L}_{R}  = \prod^{2021}_{t=1950} \text{Normal}(\varepsilon^{R}_{t}| 0, \sigma^{2}_{R}) \end{aligned}$ \\
  Fishing mortality deviation & $\begin{aligned} \mathcal{L}_{F}  = \prod^{2021}_{t=1950}  \text{Normal}(\varepsilon^{F}_{t} |0, \sigma^{2}_{F}) \end{aligned}$ \\
  Catchability deviation & $\begin{aligned} \mathcal{L}_{q}  = \prod^{2019}_{t=1977} \text{Normal}(\varepsilon^{q}_{t} | 0, \sigma^{2}_{q} ) \end{aligned}$ \\
  Selectivity deviation & $\begin{aligned} \mathcal{L}_{S}  = \text{MVN}(\boldsymbol{\varepsilon}^{S} | \boldsymbol{0}, \boldsymbol{\varSigma}) \end{aligned}$ \\
   \bf{Likelihood (observation error)} \\
CPUE & $\begin{aligned} \mathcal{L}_{I}=\prod^{2019}_{t=1976} \text{Normal}(\log(I_{t}) | \log(\widehat{I}_{t}), \sigma^{2}_{I} ) \end{aligned}$ \\
Catch & $\begin{aligned}  \mathcal{L}_{Y}  = \prod^{2021}_{t=1950}\text{Normal}(\log(Y_{t}) | \log(\widehat{Y}_{t}), \sigma^{2}_{Y} ) \end{aligned}$ \\
Length frequency & $\begin{aligned} \mathcal{L}_{LF}  = \prod^{2017}_{t=2000} \text{DM}(\boldsymbol{n}_{t} | E_{t}, \boldsymbol{\delta}_{t}) \end{aligned}$ \\
  \bf{Gamma distribution penalties} \\
CPUE error  & $\begin{aligned} f_{I} =\text{Gamma}(\sigma_{I} | 2, 10^{-10})\end{aligned}$ \\
Catch error & $\begin{aligned} f_{Y} =\text{Gamma}(\sigma_{Y} | 2, 10^{-10})\end{aligned}$ \\
Catchability deviation & $\begin{aligned} f_{q} = \text{Gamma}(\sigma_{q} | 2, 10^{-10})\end{aligned}$ \\ 
\bf{Joint likelihood for each model} & \\
M1, M2 & $\begin{aligned} \mathcal{L} (\boldsymbol{\Theta}, \boldsymbol{Q} |\boldsymbol{D}) = \mathcal{L}_{R} \cdot \mathcal{L}_{F} \cdot \mathcal{L}_{LF} \cdot \mathcal{L}_{I} \cdot \mathcal{L}_{Y} \cdot f_{I} \cdot f_{Y} \end{aligned}$ \\
M3 & $\begin{aligned} \mathcal{L} (\boldsymbol{\Theta}, \boldsymbol{Q} |\boldsymbol{D}) = \mathcal{L}_{R} \cdot \mathcal{L}_{F} \cdot \mathcal{L}_{q} \cdot \mathcal{L}_{LF} \cdot \mathcal{L}_{I} \cdot \mathcal{L}_{Y} \cdot f_{I} \cdot f_{Y} \cdot f_{q} \end{aligned}$ \\
M4, M5, M7, M8, M10, M11, M13, M14& $\begin{aligned} \mathcal{L} (\boldsymbol{\Theta}, \boldsymbol{Q} |\boldsymbol{D}) = \mathcal{L}_{R} \cdot \mathcal{L}_{F} \cdot \mathcal{L}_{S} \cdot \mathcal{L}_{LF} \cdot \mathcal{L}_{I} \cdot \mathcal{L}_{Y} \cdot f_{I} \cdot f_{Y} \end{aligned}$ \\
M6, M9, M12, M15 & $\begin{aligned} \mathcal{L} (\boldsymbol{\Theta}, \boldsymbol{Q} |\boldsymbol{D}) = \mathcal{L}_{R} \cdot \mathcal{L}_{F} \cdot \mathcal{L}_{q} \cdot \mathcal{L}_{S} \cdot \mathcal{L}_{LF} \cdot \mathcal{L}_{I} \cdot \mathcal{L}_{Y} \cdot f_{I} \cdot f_{Y} \cdot f_{q} \end{aligned}$ \\
 \hline
\end{tabular}
\end{table}

\clearpage

\begin{table}[]
\centering
\caption{Estimates of the parameters for the skew normal distribution representing length-at-age distributions in catch. $\xi_{a}$, $\omega_{a}$, and $\alpha_{a}$ are the location, scale, and shape parameters for the skew normal distribution for age $a$, respectively.} \label{table:skew_normal_catch}
\begin{tabular}{crrrrrr}
\hline   
\multirow{2}{*}{Parameter} & \multicolumn{6}{c}{Age ($a$)} \\ \cline{2-7} 
                   &  $0$    & $1$   & $2$   & $3$   & $4$   & $5$   \\
\hline                
    $\xi_{a}$      &  $21.691$ & $25.200$   & $27.471$  & $29.911$  & $39.702$  & $43.199$ \\
    $\omega_{a}$   &  $3.735$  &  $3.980$ & $4.352$     & $6.168$   & $4.400$  &  $4.132$ \\
    $\alpha_{a}$   &  $1.104$  & $1.425$  & $2.080$     & $8.687$   & $-1.016$  &  $-2.752$\\
\hline
\end{tabular}
\end{table}

\clearpage

\clearpage

\clearpage

\clearpage

\clearpage

\clearpage

\clearpage

\clearpage

\newpage
\appendix

\section{Appendix A: Initial
abundance}\label{appendix-a-initial-abundance}

The abundance at near unfished equilibrium \(N_{a,0}\) was calculated
as:

\[
\begin{aligned} 
N_{a,0}= 
\begin{cases} R, & \text{if } a = 0 \\ N_{a-1,0} \cdot e^{-Z_{a-1,0}}, & \text{if } 0 < a < A \\ \dfrac{N_{a-1,0} \cdot e^{-Z_{a-1,0}}}{1-e^{-Z_{a,0}}}, & \text{if } a=A \\ 
\end{cases}
\end{aligned}
\]

The total mortality rate at equilibrium \(Z_{a,0}\) consists of the
age-specific fishing mortality rate at equilibrium \(F_{a,0}\) and the
constant natural mortality rate \(M\) (i.e., \(Z_{a,0}=F_{a,0}+M\)).
\(F_{a,0}\) was separated into two components: the fully selected
fishing mortality rate \(F_{0}\) and the age-dependent selectivity
\(S_{a,0}\). We assumed that the fishing mortality rate during the
period before data collection was negligible and near zero, as the catch
data for the initial years of the study period showed very low values.
Therefore, the fully selected fishing mortality rate \(F_{0}\) was fixed
at 0.01, implying that \(Z_{a,0} \approx M\), and the age-dependent
selectivity \(S_{a,0}\) was calculated using Equation
\eqref{eq:SelFunction}, where \(a_{05,t}\) was replaced with \(a_{05}\).

\section{Appendix B: CDF of a skew normal
distribution}\label{appendix-b-cdf-of-a-skew-normal-distribution}

The cumulative distribution function (CDF) of a skew normal distribution
\(\Phi(\cdot)\) \citep{azzalini1985class} to model the length-at-age
distribution in catch was defined as

\[
\begin{aligned} 
\Phi(x|\xi_{a}, \omega_{a}, \alpha_{a})= \int^{x}_{-\infty} \left[ \frac{1}{\xi_{a} \cdot \pi} \cdot e^{-\frac{(L-\xi_{a})^{2}}{2 \cdot \omega^{2}} } \cdot \int_{-\infty}^{\alpha_{a} \cdot \frac{L-\xi_{a}}{\omega_{a}}} e^{-\frac{t^{2}}{2}} dt \right] dL,
\end{aligned}
\]

where \(a\) is the subscript for age, \(L\) is the length, \(x\) is the
upper limit of integration over \(L\), \(\xi_{a}\) is the location
parameter, \(\omega_{a}\) is the scale parameter, and \(\alpha_{a}\) is
the shape parameter.

\newpage

Supplementary Information

\section*{Supplementary Material}

\subsection*{Supplementary Tables}

\begin{table}[ht]
\centering
\caption{Parameter lower and upper bounds, as well as initial values of parameters and random effects, for all alternative models fitted to the Korean mackerel data.} \label{table:bounds}
\begin{tabular}{ l r r r }
\hline
 Quantity                        & Lower bound ($l$)   & Upper bound ($u$) & Initial value  \\ 
 \hline
 \bf{Parameter} \\
 $\text{logit}_{(l,u)}(a_{05})$  &  $0$            &     $2$             &  $0$ \\
 $\text{logit}_{(l,u)}(a_{95})$  &  $2$            &     $5$             &  $0$ \\
 $\text{logit}_{(l,u)}(q_{0})$   &  $10^{-7}$      &     $10^{-2}$       &  $0$ \\
 $\text{logit}_{(l,u)}(c)$       &  $10^{-8}$      &     $10^{-2}$       &  $0$ \\
 $\log(R)$                       &  \textemdash             & \textemdash                     &  $16$ \\
 $\log(\sigma_{R})$              &  \textemdash             & \textemdash             &   $0$ \\
 $\log(\sigma_{I})$              &  \textemdash             & \textemdash              &   $0$ \\
 $\log(\sigma_{F})$              &  \textemdash             & \textemdash                &   $0$ \\
 $\log(\sigma_{Y})$              &  \textemdash             & \textemdash                &   $0$ \\
 $\log(\sigma_{q})$              &  \textemdash             & \textemdash                &   $0$ \\
 $\log(\theta)$                  &  \textemdash             & \textemdash                &   $0$ \\
 \bf{Random effect} \\
 $\boldsymbol{\varepsilon}^{R}$  &  \textemdash             & \textemdash    &   $ \boldsymbol{0}$ \\
 $\boldsymbol{\varepsilon}^{S}$  &  \textemdash             & \textemdash    &   $ \boldsymbol{0}$ \\
 $\boldsymbol{\varepsilon}^{F}$  &  \textemdash             & \textemdash    &   $ \boldsymbol{0}$ \\
 $\boldsymbol{\varepsilon}^{q}$  &  \textemdash             & \textemdash    &   $ \boldsymbol{0}$ \\
 \hline
\end{tabular}
\end{table}

\clearpage

\renewcommand{\arraystretch}{1.2}
\begin{table}[]
\centering
\caption{Cross-test evaluation of AIC accuracy in model selection through 500 simulation-estimation runs for each combination of an operating Model (OM) and estimation model (EM). Rows indicate the OMs that simulated data, using parameter estimates from fitting the mackerel data as the true values. Columns represent the EMs applied to the simulated data. The table shows the frequency at which each EM was identified as the best model based on AIC. Numbers outside parentheses are from the cross-test where M12 imposed a gamma distribution penalty on the standard deviation parameter of the random walk catchability coefficient, denoted as $\sigma_{q}$. Numbers in parentheses represent outcomes where M12 did not apply the gamma distribution penalty to $\sigma_{q}$.}
\label{table:cross-test}
\begin{tabular}{rrrrr}
\hline   
\multirow{2}{*}{OM} & \multicolumn{3}{c}{EM} & \multirow{2}{*}{Total} \\ \cline{2-4} 
                   &  M10    & M11   & M12   &    \\
 \hline
 M10   & 400 (402) &  50 (50) &  0 (1)  &    450 (453)  \\
 M11   & 0 (0)  & 468 (466)   &  0 (1)  &    468 (467)  \\
 M12   & 167 (16) & 186 (33) &  1 (306) &    354 (355)  \\
 \hline
\end{tabular}
\end{table}

\clearpage

\renewcommand{\arraystretch}{1.1}
\begin{sidewaystable}[ph!]
\centering
\caption{Coverage probabilities of the 95\% confidence intervals for the fixed-effect parameters in all alternative models, based on 1000 runs of parametric bootstrap. The intervals were computed using the Wald interval.} \label{table:coverage}
\begin{tabular}{ l l l l l l l l l l l l l l l l }
\hline
 Parameter                                 &  M1 & M2 & M3 & M4 & M5 & M6 & M7 & M8 & M9 & M10 & M11 & M12 & M13 & M14 & M15  \\ 
 \hline
 $\log(R)$                                 &  95 & 91 & 93 & 94 & 93 & 92 & 93 & 93 & 92 & 93 & 93 & 93 & 94 & 91 & 93 \\
 $\text{logit}_{(0,2)}(a_{05})$            &  96 & 95 & 94 & 93 & 93 & 93 & 92 & 90 & 92 & 91 & 92 & 93 & 82 & 80 & 81 \\ 
 $\text{logit}_{(2,5)}(a_{95})$            &  95 & 94 & 95 & 95 & 95 & 95 & 96 & 95 & 96 & 95 & 95 & 96 & 95 & 95 & 96 \\
 $\log(\sigma_{Y})$                        &  92 & 92 & 91 & 92 & 92 & 88 & 90 & 90 & 87 & 91 & 92 & 89 & 89 & 93 & 88 \\
 $\log(\sigma_{I})$                        &  95 & 94 & 89 & 96 & 95 & 90 & 94 & 96 & 93 & 95 & 94 & 94 & 95 & 94 & 93 \\
 $\text{logit}_{(10^{-7},10^{-2})}(q_{0})$ &  97 & 91 & 93 & 94 & 93 & 92 & 93 & 92 & 93 & 94 & 92 & 93 & 95 & 93 & 93 \\
 $\log(\sigma_{F})$                        &  94 & 93 & 95 & 94 & 95 & 94 & 94 & 93 & 95 & 93 & 94 & 96 & 93 & 95 & 94 \\
 $\log(\sigma_{R})$                        &  95 & 95 & 95 & 96 & 94 & 98 & 94 & 94 & 98 & 96 & 96 & 97 & 97 & 97 & 97 \\
 $\log(\theta)$                            &  94 & 94 & 94 & 96 & 95 & 96 & 95 & 95 & 95 & 95 & 96 & 95 & 96 & 95 & 95 \\
 $\text{logit}_{(10^{-8},10^{-2})}(c)$     &  \textemdash & 91 & \textemdash & \textemdash & 92 & \textemdash & \textemdash & 93 & \textemdash & \textemdash & 94 & \textemdash & \textemdash & 92 & \textemdash \\
 $\log(\sigma_{q})$                        &  \textemdash & \textemdash & 95 & \textemdash & \textemdash & 98 & \textemdash & \textemdash & 97 & \textemdash & \textemdash & 96 & \textemdash & \textemdash & 96 \\
 $\log(\sigma_{S})$                        &  \textemdash & \textemdash & \textemdash & 94 & 93 & 92 & 95 & 94 & 92 & 96 & 95 & 93 & 93 & 93 & 93 \\
 \hline
\end{tabular}
\end{sidewaystable}

\clearpage

\subsection*{Supplementary Figures}

\begin{figure}

{\centering \includegraphics[width=0.95\linewidth]{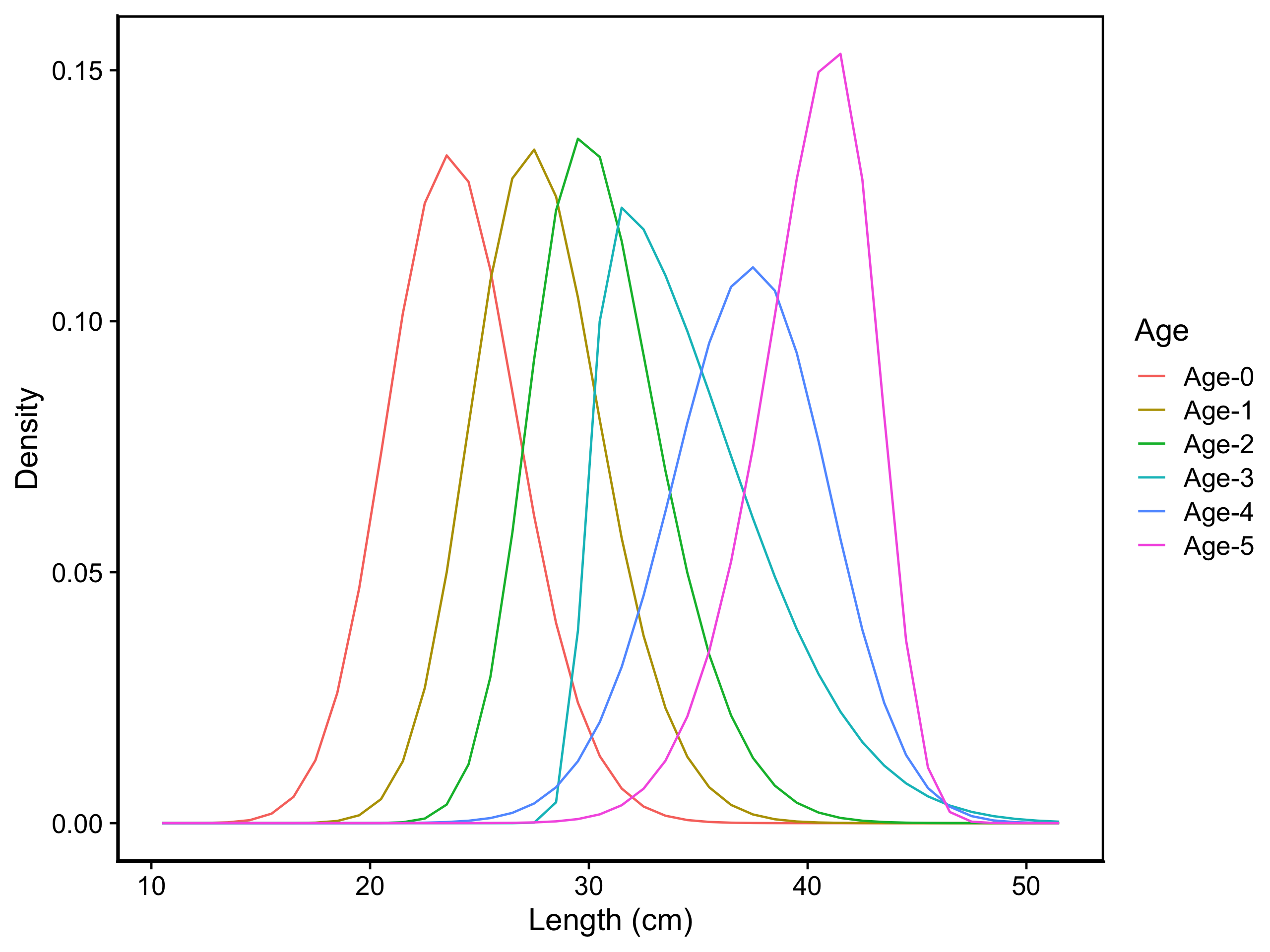} 

}

\caption{Length-at-age distributions estimated using the minimum, mean, and maximum lengths at age from \citet{jung2021study}. The minimum and maximum lengths at age were employed to match the 5\% and 95\% percentiles of the skew normal distribution, respectively, while the mean length-at-ages were used to match the mean of each distribution. \label{LengthAge}}\label{fig:unnamed-chunk-35}
\end{figure}

\clearpage

\begin{figure}

{\centering \includegraphics[width=0.95\linewidth]{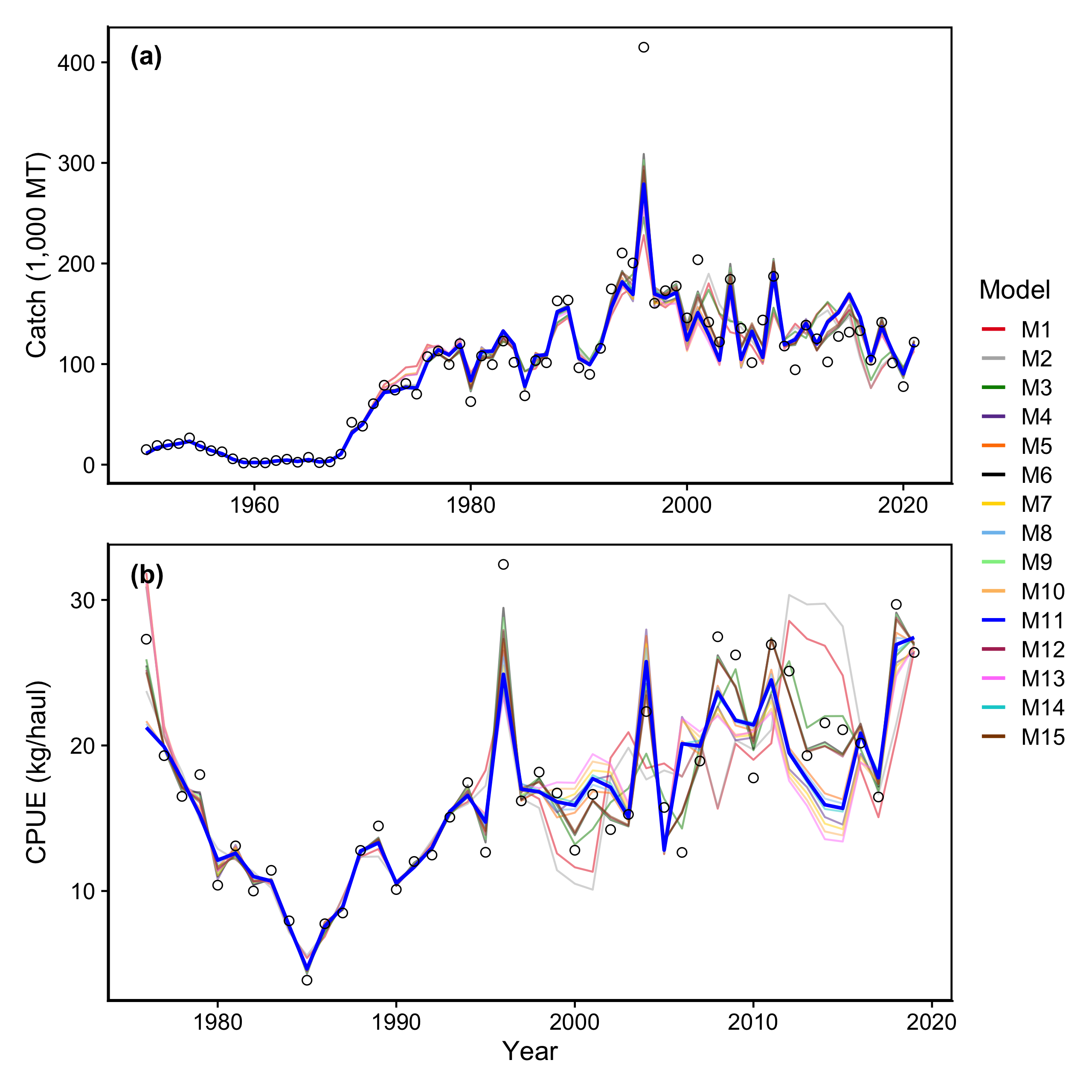} 

}

\caption{Model-fitted values (solid lines) from all 15 alternative models and observed data (points) for the catch (a) and catch-per-unit-effor (CPUE) (b) data. Each color of the line represents a different model, and the blue line represents the best model (M11). \label{FittedValues}}\label{fig:unnamed-chunk-36}
\end{figure}

\clearpage

\begin{figure}

{\centering \includegraphics[width=0.95\linewidth]{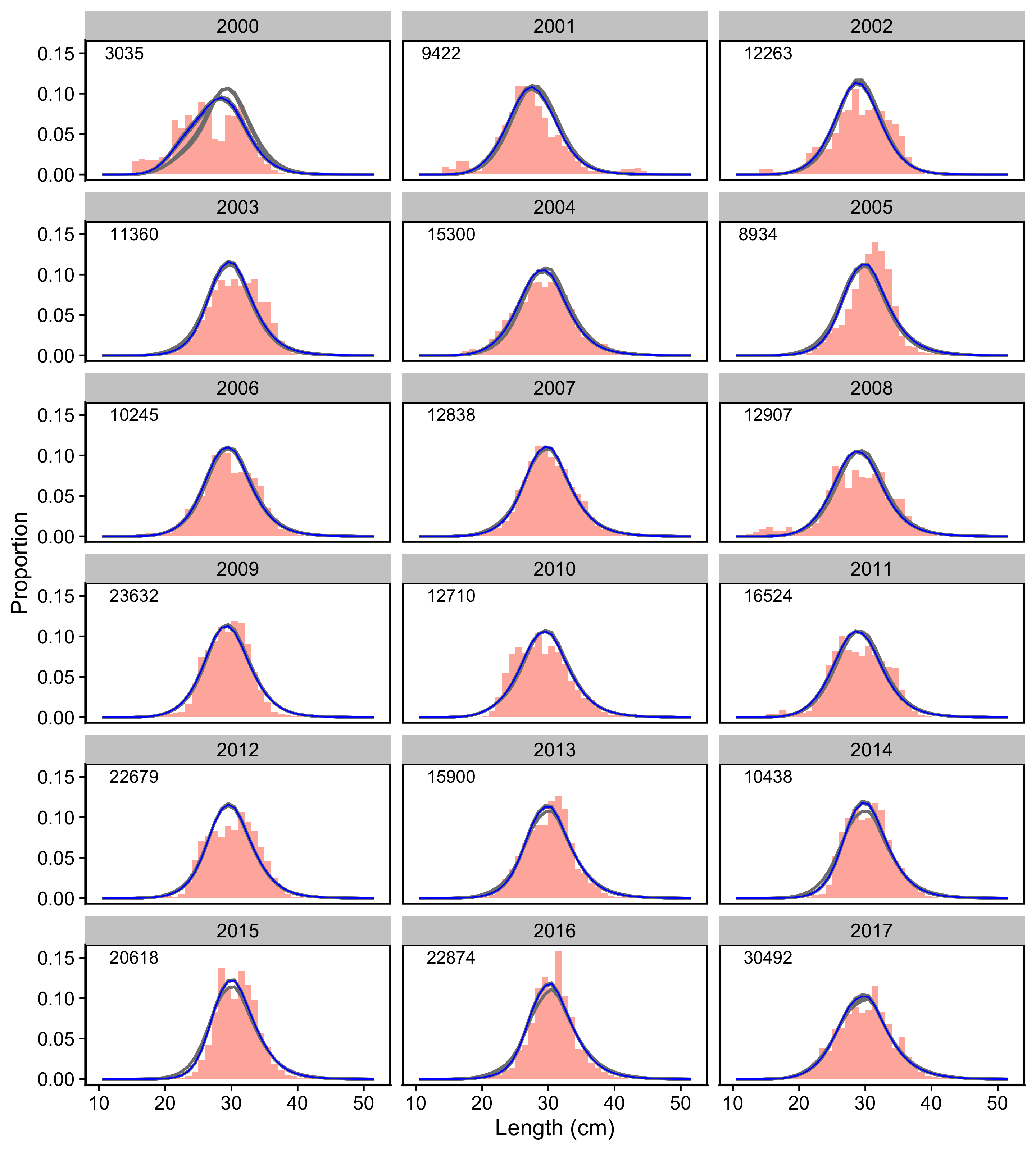} 

}

\caption{Model-fitted values (solid lines) from all 15 alternative models and observed data (histograms) for the length composition data in proportion. Each line represents a different model, and the blue line represents the best model (M11). Each panel corresponds to a specific year, indicated in the top of each panel. The number of observations for each year is indicated in the top-left corner of each panel. \label{FittedValuesLF}}\label{fig:unnamed-chunk-37}
\end{figure}

\clearpage

\begin{figure}

{\centering \includegraphics[width=0.95\linewidth]{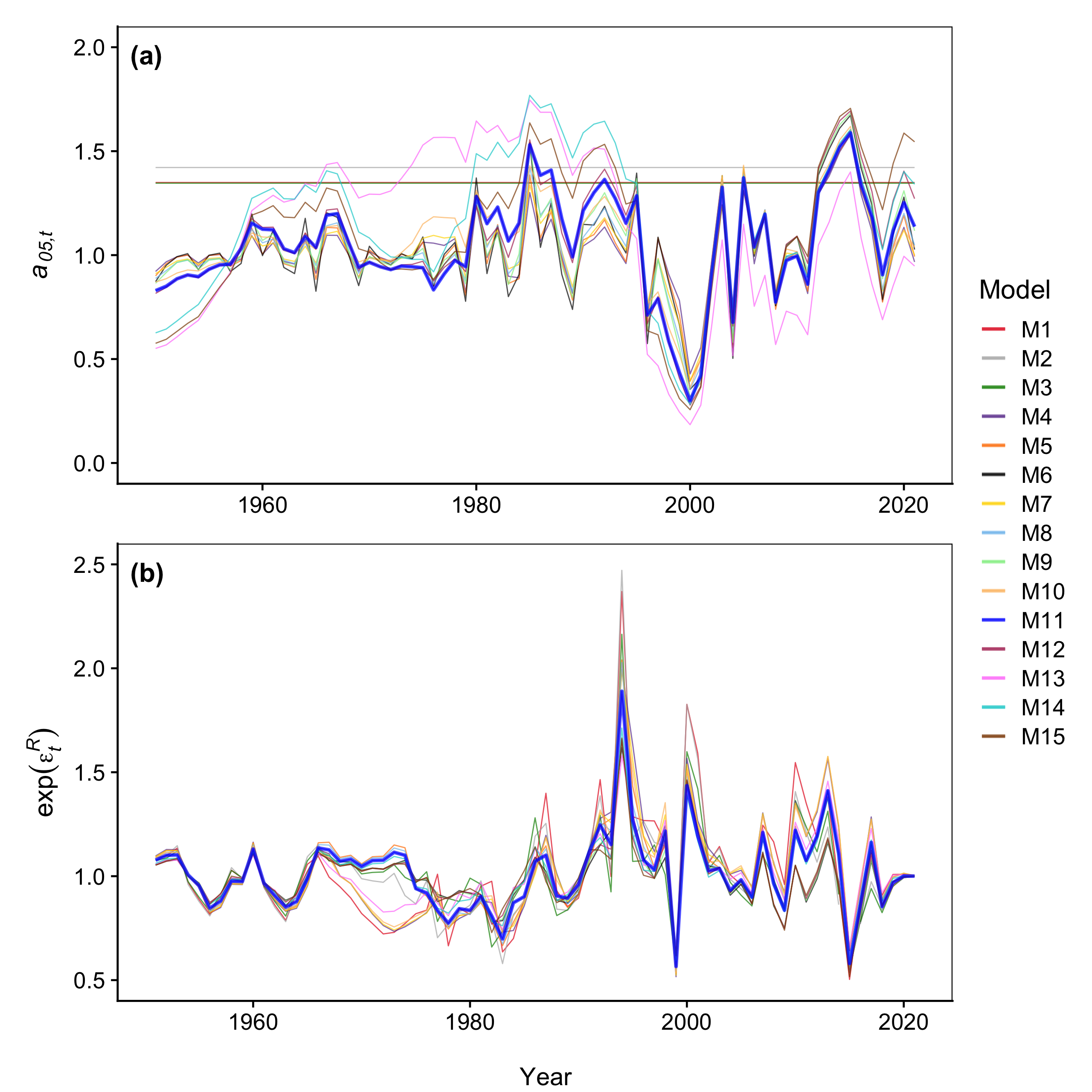} 

}

\caption{Estimates of the time-varying age at 5\% selectivity ($a_{05, t}$) (a) and those of the exponent of recruitment deviations ($\exp(\varepsilon^{R}_{t})$) (b) from all 15 alternative models. Each color of the line represents a different model, and the blue line represents the best model (M11). \label{SelRecEstim}}\label{fig:unnamed-chunk-38}
\end{figure}

\clearpage

\clearpage

\begin{figure}

{\centering \includegraphics[width=0.95\linewidth]{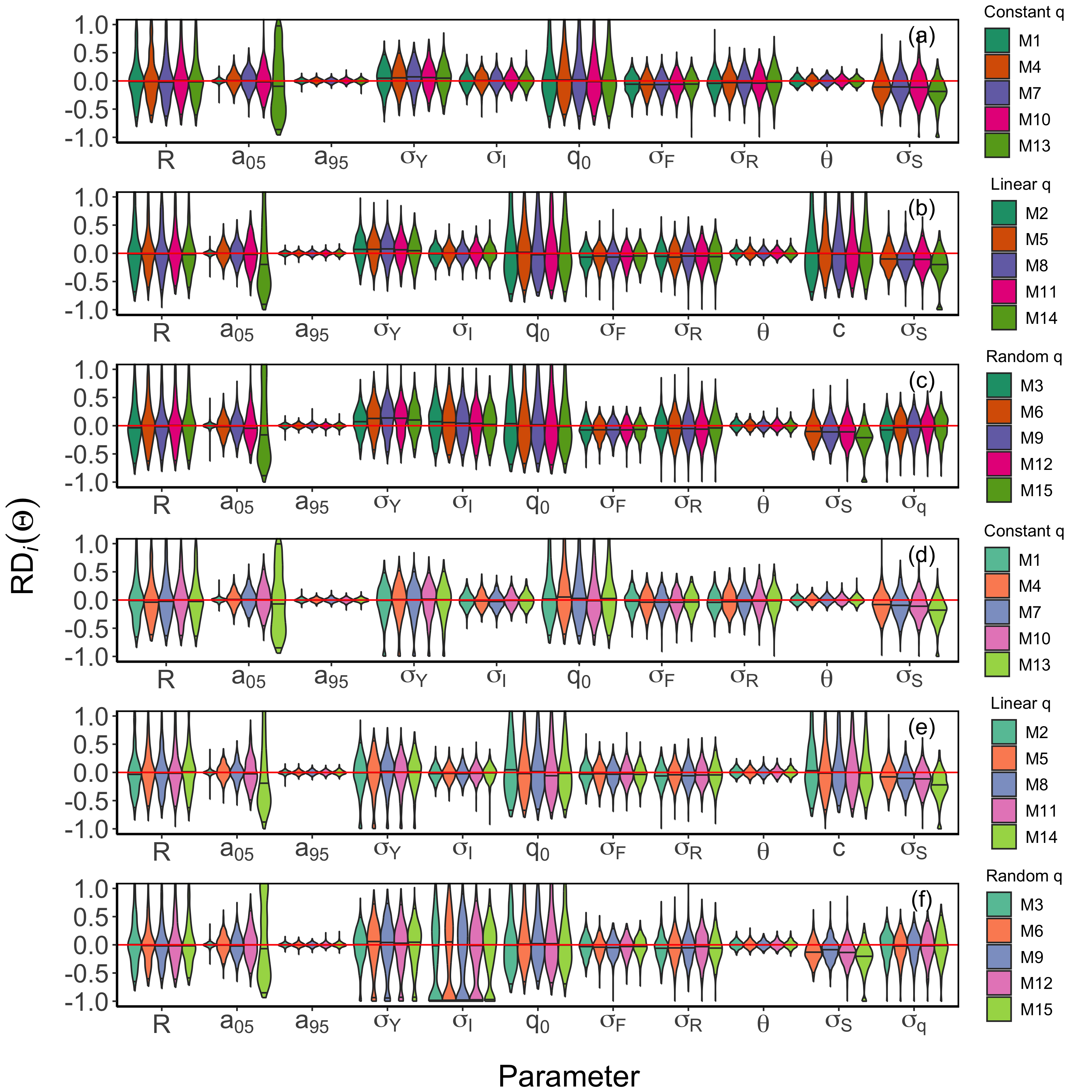} 

}

\caption{Relative difference (RD) of the estimates of the fixed effect parameters for all alternative models (M1-15). The horizontal red line indicates no difference (RD = 0) between the estimates and their corresponding true values. The violin plots in each panel show the distributions of RD values from models under three catchability assumptions: constant catchability (a and d), linearly increasing catchability (b and e), and random walk catchability (c and f). Panels (a) to (c) display RD values for the estimates from models with gamma penalties on the standard deviation parameters, while panels (d) to (f) show RD values for estimates from models without gamma penalties on the standard deviation parameters. The horizontal lines within each violin plot indicate the 2.5\%, 50\%, and 97.5\% percentiles of the RD values. \label{MackBoot}}\label{fig:unnamed-chunk-40}
\end{figure}

\clearpage

\begin{figure}

{\centering \includegraphics[width=0.95\linewidth]{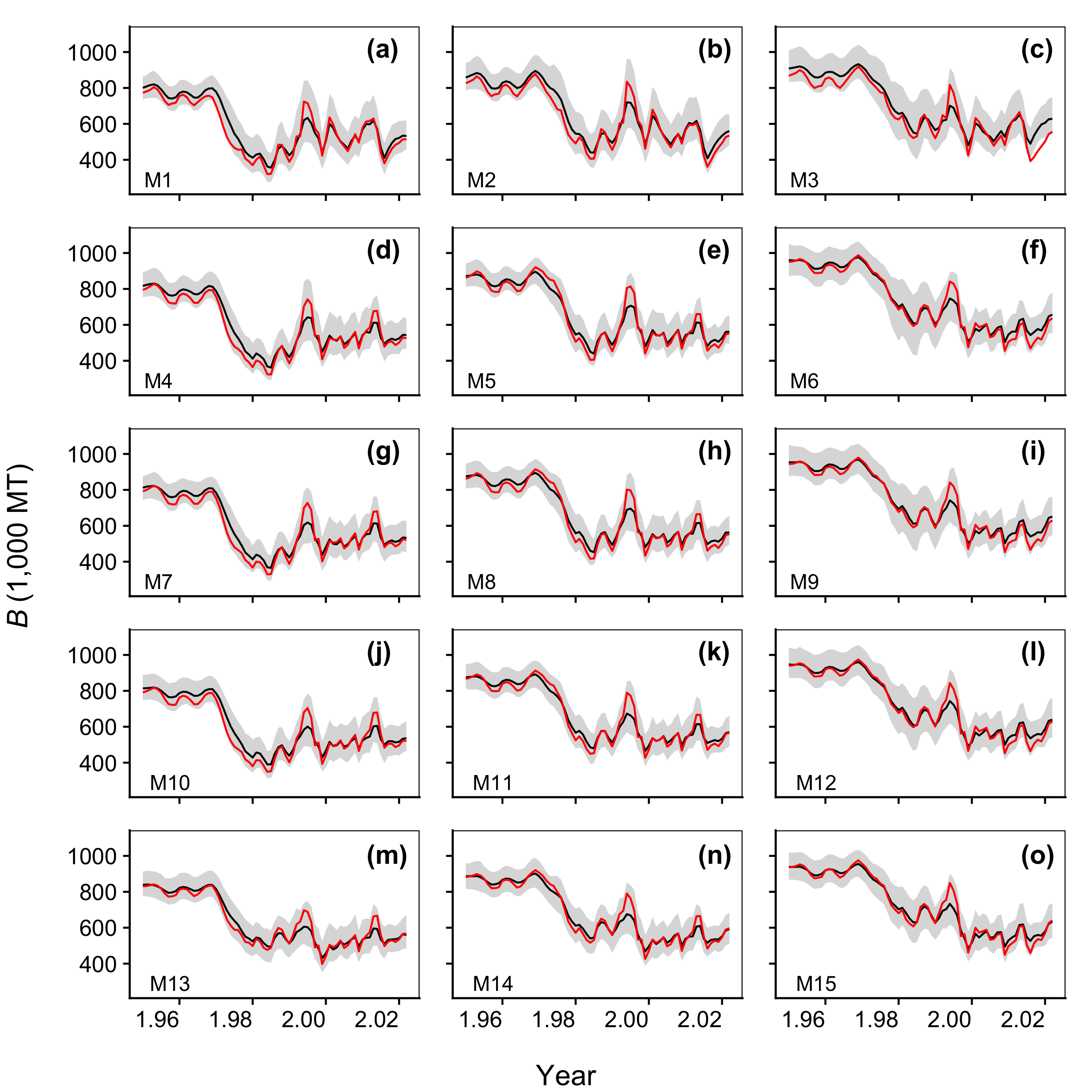} 

}

\caption{Self-test estimates of the total biomass ($B$) and corresponding true values for all alternative models (M1-15), with the name of the corresponding model indicated in the bottom-left corner of each panel. The black lines indicate the median values of the self-test estimates, and the gray areas indicate the 95\% intervals. The red lines indicate the true values. \label{MackSelfB}}\label{fig:unnamed-chunk-41}
\end{figure}

\clearpage

\clearpage

\clearpage

\clearpage

\clearpage

\clearpage

\end{document}